\documentclass[11pt]{article}
\usepackage{graphicx}

\usepackage{txfonts}
\usepackage{graphicx}
\usepackage{multicol}
\usepackage{indentfirst}
\usepackage{subfigure}
\usepackage{tabularx}
\usepackage{times}
\usepackage{harpoon}
\usepackage{color}
\usepackage[superscript]{cite}
\usepackage{subfig}
\usepackage{siunitx}
\usepackage{ulem}

\usepackage{color}

\begin{document}
\sloppy
\doublerulesep=0pt \footnotesep=0pt
%--------------------------- Head and Foot Format --------------------------
\catcode`@=11
\def\@evenhead{\vbox{\hbox to\textwidth{{\sf\small\evenhead}}
    \protect\vspace{0.3truemm}\relax
    \hrule depth0pt height0.15truemm width\textwidth}}
\def\@oddhead{\vbox{\hbox to\textwidth{{\sf\small\oddhead}}
    \protect\vspace{0.3truemm}\relax
    \hrule depth0pt height0.15truemm width\textwidth}}
\def\@evenfoot{\hfil\evenfoot-\thepage\hfil}
\def\@oddfoot{\hfil\evenfoot-\thepage\hfil}
\def\@seccntformat#1{\csname the#1\endcsname. }
\renewcommand\section{
\@startsection {section}{1}{\z@}%
{-2.3ex \@plus -1ex \@minus -.2ex}%
{1.8ex \@plus.2ex}%
{\normalfont\large\bfseries}} \catcode`@=12
\setlength{\columnsep}{6truemm} \setlength{\footnotesep}{3truemm}
\def\REF#1{\par\hangindent\parindent\indent\llap{#1\enspace}\ignorespaces}
\def\SEC#1{\vspace{0.1\baselineskip}\vbox{\vspace*{1.0\baselineskip}
\noindent\large\textbf{#1}}\vspace*{0.8\baselineskip}\rm}
\renewcommand\abstractname{\vspace*{-2\baselineskip}}
%========================= Definitions =============================
\def\evenhead{ \hfil Jinou Dong \textit{et al.}}
\def\oddhead{Jinou Dong \textit{et al.} \hfil }
\def\evenfoot{}
%========================= Begin Document ============================
\begin{center}\Large\bf
Spin disorder state induced by Mg$^{2+}$ doping in a Kitaev material Na$_{3}$Co$_{2}$SbO$_{6}$
\end{center}
\begin{center}\large\rm
Jinou Dong$^{1}$, Xueqin Zhao$^{1}$, Lingfeng Xie$^{1}$, Xun Pan$^{1}$, Haoyuan Tang$^{1}$, Zhicheng Xu$^{1}$, Guoxiang Zhi$^{2}$, Chao Cao$^{1, 3}$,  Xiaoqun Wang$^{1, 3\dag}$ and
Fanlong Ning $^{1,3, 4, 5\dag}$
\\[2
mm] {\small\it  $^1$ School of Physics, Zhejiang University, \\Hangzhou 310027, China\\
$^2$ Tianmushan Laboratory,\\Hangzhou 310023, China\\
$^3$ Institute for Advanced Study in Physics, Zhejiang University, \\Hangzhou, 310058, China\\
$^4$ State Key Laboratory of Silicon and Advanced Semiconductor Materials, Zhejiang University, \\Hangzhou 310027, China\\
$^5$ Collaborative Innovation Center of Advanced Microstructures, Nanjing University, \\Nanjing 210093, China\\}
\end{center}

\renewcommand{\footnoterule}{\flushleft\rule{4cm}{0.5pt}\vspace{-0.5pt}}
\footnotesep=5mm\footnotetext {$\!\!\!\!\!\dag$ \small Corresponding
author. Email: ningfl@zju.edu.cn, xiaoqunwang@zju.edu.cn
\hfill\par

%\small Received  \hfill \copyright \,\,
%Institute of Electronics

 \hfill\par \vspace*{1mm}
\centerline{\normalsize\evenfoot-\thepage}
\par \vspace*{-12.4mm}\par}

\begin{abstract}\normalsize{
\noindent\textbf{Abstract:}\,\,\,Due to the dominant Kitaev exchange interactions, 
the cobaltate, Na$_{3}$Co$_{2}$SbO$_{6}$, has been considered to be approximate to the Kitaev quantum spin liquid (QSL). 
Here, we investigate both magnetic dilution and chemical pressure effects of Na$_{3}$Co$_{2}$SbO$_{6}$ by the substitutions of Mg$^{2+}$ for Co$^{2+}$ through the structural, optical, magnetic and thermodynamic measurements.
No structural transition has been observed, and the bandgaps remain constant in all doping levels.
Combining with the magnetic and thermodynamic measurements, 
we find that the antiferromagnetic transition temperature is continuously 
suppressed with increasing Mg doping levels and completely disappears at $x=0.2$.
Interestingly, when the doping level $x$ is larger than 0.2, neither long-range magnetic order nor spin glass state has been detected, and the specific heat has a residual linear term at zero field. All features indicate that Na$_{3}$(Co$_{2-x}$Mg$_{x}$)SbO$_{6}$ system enters into a novel spin disorder (NSD) state. 

\vspace{0.5\baselineskip}

\noindent\textbf{Key words:}\,\,\,Quantum spin liquids, Kitaev model, 
Doping, Na$_{3}$Co$_{2}$SbO$_{6}$

%\noindent\textbf{DOI:}\,\,\,10.1088/1674-4926/30/1/014003\hspace{10mm}\textbf{PACC:}\,\,\,7220J£»7340Q
}
\end{abstract}

%\tableofcontents
\section{Introduction}
Quantum spin liquids (QSLs) have attracted great attentions due to the novel ground states with highly entangled spins and the absence of the long-range magnetic order down to zero temperature, which, in some cases,  can be characterized by quantum number 
fractionalization and gapless excitations without symmetry breaking. \cite{savary2016quantum, li2015rare, balents2010spin, zhou2017quantum}
In the triangular lattice, spin-exchange interactions between different lattice 
sites compete because of geometry frustration, and usually no static magnetic 
order is formed in the ground state. In 1973, Anderson constructed the resonating-valence-bond (RVB) theory, which has drawn a lot of interests in the antiferromagnetic triangular 
lattice with $S$ = 1/2. \cite{anderson1973resonating}
While RVB theory can not provide an exact solution, and no consensus has been reached on the nature of QSLs. As an alternative option, Kitaev proposed 
an exactly solvable QSL model on the honeycomb lattice with $S$ = 1/2. \cite{kitaev2006anyons} In 
Kitaev model, due to the bond-dependent anisotropic interactions, i.e., Kitaev interactions, strong quantum 
fluctuations rise, which leads to a magnetically disordered state. It has been proven that some systems may possess QSL ground states, which enlighten us for further investigation of them. \cite{lyu2022tunneling, kitaev2006anyons, winter2017models, takagi2019concept, zhou2017quantum, broholm2020quantum, trebst2022kitaev}

Following the theory proposed by Jackeli and Khaliullin, \cite{jackeli2009mott} 
many researches have been focused on the strong spin-orbit-coupled heavy d$^{5}$ transition-metal Mott 
insulators with honeycomb structure. \cite{takagi2019concept, trebst2022kitaev, winter2017models, jackeli2009mott, rau2014generic, motome2020hunting}
Among these materials, the 
first Kitaev material 5d$^{5}$ Na$_{2}$IrO$_{3}$ \cite{hwan2015direct} and 4d$^{5}$ ${\alpha}$-RuCl$_{3}$ are the representative candidates.
Due to the observed antiferromagnetic transition at low temperature 
($T_N$ = 15 K for Na$_{2}$IrO$_{3}$ and $T_N$ = 7 K for ${\alpha}$-RuCl$_{3}$), \cite{singh2012relevance, sears2015magnetic} 
they do not exhibit the desirable QSL ground state. However, 
there are mounting evidences that these systems are proximate to the Kitaev QSL states. \cite{takagi2019concept, trebst2022kitaev, hwan2015direct, choi2019spin, singh2012relevance, banerjee2016proximate, banerjee2017neutron, do2017majorana}
In particular, the zig-zag order at low temperature may be destroyed in ${\alpha}$-RuCl$_{3}$ by a high magnetic field within the honeycomb plane, and a field-induced QSL can be observed. \cite{zhao2022neutron, kim2020proximate, li2021identification, zheng2017gapless}
Nevertheless, the nature of a magnetically disordered state driven by the external field is still an open question. \cite{yu2018ultralow, hentrich2018unusual, kasahara2018unusual}
More recently, the Kitaev 
model has been extended to 3d transition-metal materials, and the 3d$^{7}$ cobaltates with a high-spin electronic configuration ($t_{2g}^{5}$$e_{g}^{2}$) have attracted a 
lot of attentions. 
It has been proposed that Co$^{2+}$ ions under an octahedral crystal field of oxygen can give 
rise to pseudospin $J_{eff}$ = 1/2, which can realize Kitaev interactions. \cite{liu2018pseudospin} In d$^{7}$ cobalt compounds, $t_{2g}$ states 
can be described by the effective orbital moment $L_{eff}$ = 1.
Combining with spin $S$ = 3/2, they can lead to the total moment $J_{eff}$ = 1/2. 
Interestingly, in d$^{7}$ system, the antiferromagnetic interactions from $t_{2g}$ electrons can be compensated by the ferromagnetic interactions from e$_{g}$ spins. \cite{liu2018pseudospin, sano2018kitaev, liu2020kitaev, motome2020materials, vavilova2023magnetic} It can tune the relative magnitude
of Heisenberg and Kitaev interactions, and drive the system into a QSL state. 
With these features, honeycomb-layered magnets Na$_{2}$Co$_{2}$TeO$_{6}$ and Na$_{3}$Co$_{2}$SbO$_{6}$ attract plenty of attentions. \cite{viciu2007structure, berthelot2012studies}
Similar to ${\alpha}$-RuCl$_{3}$, such systems also undergo antiferromagnetic 
transitions at low temperature ($T_N$ = 27 K for Na$_{2}$Co$_{2}$TeO$_{6}$ and $T_N$ = 8 K for 
Na$_{3}$Co$_{2}$SbO$_{6}$, respectively). \cite{bera2017spinon, wong2016zig, chamorro2020chemistry, wen2019experimental, kim2021spin}
It is natural to wonder what would 
happen if the magnetic order is broken in these systems, considering the fragile zig-zag order which is caused by the competition between Kitaev and non-Kitaev interactions. Previously, high magnetic field has been applied, with a temptation of transforming these systems into a proximate spin liquid state, or even directly into 
a pure Kitaev QSL state. \cite{lin2021field, yao2020ferrimagnetism, hong2021strongly, vavilova2023magnetic}
On the other hand, doping is an alternative path to explore the nature of Kitaev materials, leading the dilution system to be a platform to investigate a possible QSL state. \cite{dantas2022disorder, kao2021vacancy, do2020randomly}
It has been demonstrated that the ground state in both Na$_{2}$Co$_{2}$TeO$_{6}$ and Na$_{3}$Co$_{2}$SbO$_{6}$ can be tuned by doping Zn$^{2+}$ ions. \cite{fu2023suppression, fu2023signatures}

In this paper, we report the successful substitutions of Mg$^{2+}$ for Co$^{2+}$ in Na$_{3}$Co$_{2}$SbO$_{6}$, where the ionic radius of Mg$^{2+}$ (0.65Å) is almost 14$\%$ smaller than Co$^{2+}$ (0.74Å).
We investigate both magnetic dilution and chemical pressure effects through the structural, optical, magnetic and thermodynamic 
measurements. For all high-quality polycrystalline Na$_{3}$(Co$_{2-x}$Mg$_{x}$)SbO$_{6}$, no structural transition has been observed, and the bandgaps remain constant in all doping levels. 
Magnetic susceptibility and specific heat 
results show the gradual suppression of antiferromagnetic order with increasing Mg 
doping, which completely disappears at $x = 0.2$. Interestingly, when the doping level is high enough, neither signal of long-range magnetic order nor spin freezing has been observed in Na$_{3}$(Co$_{2-x}$Mg$_{x}$)SbO$_{6}$ system, and the magnetic specific heat can be fitted by a linearly temperature-dependent term at zero field. All these features indicate that Na$_{3}$(Co$_{2-x}$Mg$_{x}$)SbO$_{6}$ system eventually enters into a novel spin disorder (NSD) state with enough magnetic vacancies. 

\section{Experiments}
We synthesized the Na$_{3}$(Co$_{2-x}$Mg$_{x}$)SbO$_{6}$ ($x = 0, 0.05, 0.1, 0.15, 0.2, 0.3, 0.4$) and nonmagnetic 
Na$_{3}$Mg$_{2}$SbO$_{6}$ polycrystalline specimens by solid-state reaction with 
the high-purity materials of Na$_{2}$CO$_{3}$ (99.997$\%$), Co$_{3}$O$_{4}$ (99.9985$\%$), 
MgO (99.99$\%$), and Sb$_{2}$O$_{3}$ (99.999$\%$). According to the chemical formula, 
we mixed the raw materials in an evacuated silica tube, whereas 10$\%$ 
excess Na$_{2}$CO$_{3}$ was added to compensate for the loss due to
volatilization upon heating. These materials were slowly heated up to 
{800}$^{\circ}$C in {20}{h}, and the mixture was held for about 
{48}{h} before cooling down to the room temperature.  The products 
were then grounded, pelleted, and sintered at {900}$^{\circ}$C for another 
{48}{h} again to achieve the complete reaction.

The resulted samples were characterized by PANalytical x-ray diffractometer 
(model EMPYREAN) with monochromatic Cu $K_{\alpha1}$ radiation. Raman measurement 
was performed at 300 K by using the Renishaw microspectrometer equipped with a 
532 nm solid-state laser. The UV-Vis-NIR optical diffuse
reflectance spectra were obtained at the room temperature
on an Agilent Cary 5000 spectrometer using a BaSO$_{4}$ plate
as the standard (100$\%$ reflectance). The DC magnetic properties of the polycrystals were examined by using the Quantum 
Design magnetic property measurement system (MPMS-3). The AC susceptibility 
and specific heat were measured on sintered pellets by the Quantum Design physical 
property measurement system (PPMS).

\section{Measurement and Results}
\subsection{Crystal structure and lattice parameters}
\begin{figure}[htbp]
	\centering
	\includegraphics[width=0.9\textwidth]{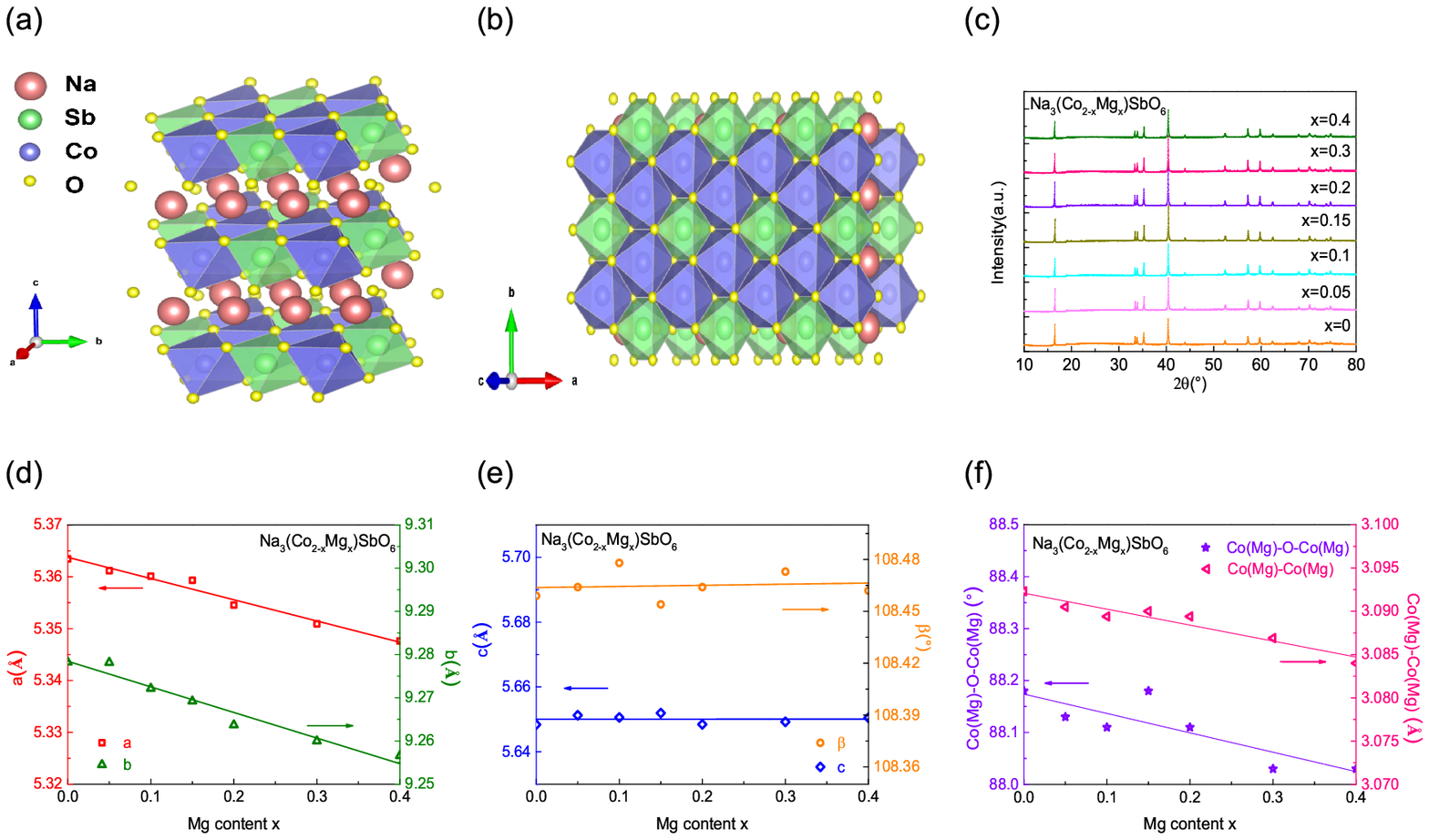}
	\caption{(a) The schematic crystal structure of Na$_{3}$Co$_{2}$SbO$_{6}$. (b) The honeycomb layer structrure of Na$_{3}$Co$_{2}$SbO$_{6}$.
		(c) The x-ray diffraction patterns of Na$_{3}$(Co$_{2-x}$Mg$_{x}$)SbO$_{6}$ ($x = 0, 0.05, 0.1, 0.15, 0.2, 0.3, 0.4$). 
		(d) The lattice parameters $\textit{a}$ and $\textit{b}$ of Na$_{3}$(Co$_{2-x}$Mg$_{x}$)SbO$_{6}$ ($x = 0, 0.05, 0.1, 0.15, 0.2, 0.3, 0.4$).
		(e) The lattice parameters $\textit{c}$ and $\beta$ of Na$_{3}$(Co$_{2-x}$Mg$_{x}$)SbO$_{6}$ ($x = 0, 0.05, 0.1, 0.15, 0.2, 0.3, 0.4$). 
		(f) The bond length of Co(Mg)-Co(Mg) and the bond-angle of Co(Mg)-O-Co(Mg) of Na$_{3}$(Co$_{2-x}$Mg$_{x}$)SbO$_{6}$ ($x = 0, 0.05, 0.1, 0.15, 0.2, 0.3, 0.4$). 
		The solid lines in (d), (e) and (f) are guides to the eye.}
	\label{f1}
\end{figure}

In Fig.\ref{f1}(a) and (b), we show the schematic crystal structure 
and the honeycomb layer structure of Na$_{3}$Co$_{2}$SbO$_{6}$.
Six edge-shared CoO$_{6}$ octahedra and one centrally 
located SbO$_{6}$ octahedra form a magnetic honeycomb layer that is 
arranged along the \textit{c}-axis. The intermediate nonmagnetic Na$^{+}$ ions separate 
the honeycomb layers, and blocks the Co-Co super-exchange interaction 
along the \textit{c}-axis. It results in the magnetic coupling within the ab plane, which forms 
a quasi-two-dimensional magnetic structure. \cite{yan2019magnetic}
In Fig.\ref{f1}(c), we show the XRD patterns of Na$_{3}$(Co$_{2-x}$Mg$_{x}$)SbO$_{6}$ 
($x = 0, 0.05, 0.1, 0.15, 0.2, 0.3, 0.4$). 
The Bragg peaks can be well indexed by a monoclinic crystal structure with 
$C2/m$ space group for all doping samples, which is the same as 
${\alpha}$-RuCl$_{3}$. \cite{johnson2015monoclinic, cao2016low}. Moreover, no peak splitting or 
additional peak has been observed, indicating no structural transition in all doping compounds.
For the parent compound Na$_{3}$Co$_{2}$SbO$_{6}$, the lattice parameters 
can be obtained through Rietveld refinement by using the GSAS-II package, \cite{toby2013gsas} and \textit{a}=5.3634(4)(Å), \textit{b}=9.2782(2)(Å), \textit{c}=5.6483(9)(Å) and ${\beta}$=108.45(9)$^{\circ}$, 
respectively. These results are consistent with previous works. \cite{fu2023signatures, wong2016zig, yan2019magnetic, viciu2007structure}
Additionally, the refinement parameters are 
$R_{wp}$ $\approx$ 3.65$\%$ and $\chi^{2}$ $\approx$ 1.72, implying a well and reliable refinement.
Following the same refinement method, we obtained the lattice parameters of all doping samples and show them
in Fig.\ref{f1}(d), (e) and (f).
We note that \textit{a} and \textit{b} are continuously decreasing, 
while \textit{c} and $\beta$ remain almost constants with increasing Mg contents.
Accordingly, the ratio of 
\textit{c} to \textit{a} is increasing with increasing doping.
Moreover, the obtained bond length of Co(Mg)-Co(Mg) and bond-angle of Co(Mg)-O-Co(Mg) show the decreasing trends with increasing Mg content $x$.
In other words, doping Mg is equivalent to the application of positive chemical pressure in $ab$ plane, and the 
quasi-two-dimensional property of the system is enhanced via Mg doping.
The monotonic behaviors of changes in lattice parameters are also the indications of homogeneous substitutions of Mg$^{2+}$ for Co$^{2+}$ in 
Na$_{3}$Co$_{2}$SbO$_{6}$.

\begin{figure}[htbp]
	\centering
	\includegraphics[width=0.9\textwidth]{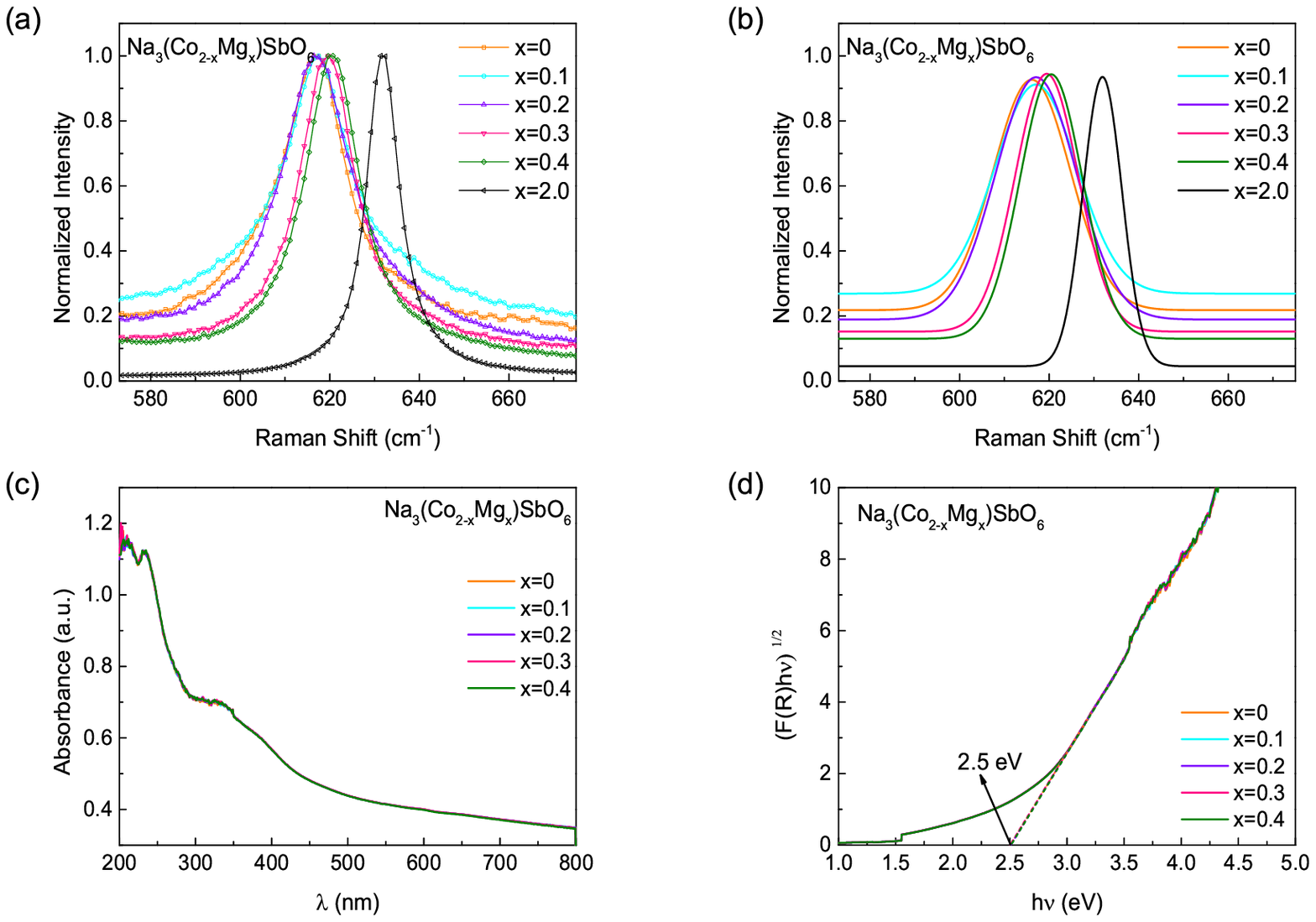}
	\caption{(a) Raman spectra of Na$_{3}$(Co$_{2-x}$Mg$_{x}$)SbO$_{6}$ ($x = 0, 0.1, 0.2, 0.3, 0.4$) and Na$_{3}$Mg$_{2}$SbO$_{6}$ measured at the room temperature. 
		(b) Gaussian fitting of Raman spectra of Na$_{3}$(Co$_{2-x}$Mg$_{x}$)SbO$_{6}$ ($x = 0, 0.1, 0.2, 0.3, 0.4$) and Na$_{3}$Mg$_{2}$SbO$_{6}$. 
		(c)Absorbance spectra transformed by diffuse reflectance spectra versus wavelength for Na$_{3}$(Co$_{2-x}$Mg$_{x}$)SbO$_{6}$ ($x = 0, 0.1, 0.2, 0.3, 0.4$). %The dash lines indicate the bandgaps.
		(d) Tauc plot of Na$_{3}$(Co$_{2-x}$Mg$_{x}$)SbO$_{6}$ ($x = 0, 0.1, 0.2, 0.3, 0.4$); the intersections of the dash lines and the \textit{x}-axis indicate the values of bandgaps; the small discontinuity near 1.5 eV is the result of a lamp change, and will not affect the bandgap calculation.}
	\label{f2}
\end{figure}

To further verify the microscopic homogeneity, we measure the Raman spectroscopy of Na$_{3}$(Co$_{2-x}$Mg$_{x}$)SbO$_{6}$ ($x = 0, 0.1, 0.2, 0.3, 0.4$)  and the end compound Na$_{3}$Mg$_{2}$SbO$_{6}$, and show the results in Fig.\ref{f2}(a). 
For the parent compound Na$_{3}$Co$_{2}$SbO$_{6}$, the 
result is consistent with the previous work. \cite{ponosov2024raman} 
%It is a strong proof in the microscopic homogeneity of substitution. 
We also show the Gaussian fitting
of Raman spectra in Fig.\ref{f2}(b).
Apparently, the central position of peak ($x_{c}$) moves to high Raman shift region with increasing $x$, indicating that the bond length of Co(Mg)-Co(Mg) becomes shorter and the strength of bond becomes larger with increasing doping.
Considering the localized  
nature of the Raman vibrations, no elemental segregation has been detected, and the Mg substitutions are homogenous. Similar results have been shown in Na(Yb$_{1-x}$Lu$_{x}$)Se$_{2}$. \cite{cairns2022tracking} 
The $x_{c}$ and the full width at half maxima (FWHM) of Na$_{3}$(Co$_{2-x}$Mg$_{x}$)SbO$_{6}$ ($x = 0, 0.1, 0.2, 0.3, 0.4$) and Na$_{3}$Mg$_{2}$SbO$_{6}$ are tabulated in Table \ref{table1}. 
We plot the absorption spectra in Fig.\ref{f2}(c), which has been transformed by a Kubelka-Munk-Function \cite{kubelka1931contribution} based on the diffuse reflection spectra. We obtain the bandgaps from a Tauc-Plot, \cite{tauc1966optical} which is shown in Fig.\ref{f2}(d). 
For the parent compound Na$_{3}$Co$_{2}$SbO$_{6}$, the bandgap is estimated as 2.50(3) eV, which is roughly consistent with the reported value ($\sim$ 2.12(3) eV). \cite{li2023investigation}
We notice that the bandgaps of all Na$_{3}$(Co$_{2-x}$Mg$_{x}$)SbO$_{6}$ ($x = 0, 0.1, 0.2, 0.3, 0.4$) samples remain constant $\sim$ 2.50 eV, 
which show the similar behavior as that of Na(Yb$_{1-x}$Lu$_{x}$)S$_{2}$ series. \cite{haussler2022diluting}
The obtained bandgaps of Na$_{3}$(Co$_{2-x}$Mg$_{x}$)SbO$_{6}$ ($x = 0, 0.1, 0.2, 0.3, 0.4$) are tabulated in Table \ref{table2}.

\begin{table}[htbp]
	\centering
	%\captionsetup{font=scriptsize}
	\caption{The central position of peak ($x_{c}$) and the full width at half maxima (FWHM) of Na$_{3}$(Co$_{2-x}$Mg$_{x}$)SbO$_{6}$ ($x = 0, 0.1, 0.2, 0.3, 0.4$) and Na$_{3}$Mg$_{2}$SbO$_{6}$.}
		\begin{tabular}{ccccccc}
					\hline
			$x$      & 0  & 0.1 & 0.2 & 0.3 & 0.4 & 2.0 \\
			\hline
			$x_{c}$ (cm$^{-1}$)        & 616.11   & 617.05   & 617.08   & 619.45  & 620.46 &  631.90 \\
			FWHM (cm$^{-1}$)        & 20.30   & 21.57   & 21.21   & 16.57  & 16.43 &  10.29 \\
					\hline
		\end{tabular}
		\label{table1} 
\end{table}

\begin{table}[htbp]
	\centering
	%\captionsetup{font=scriptsize}
	\caption{The bandgaps of Na$_{3}$(Co$_{2-x}$Mg$_{x}$)SbO$_{6}$ ($x = 0, 0.1, 0.2, 0.3, 0.4$).}
		\begin{tabular}{cccccc}
					\hline
			$x$      & 0  & 0.1 & 0.2 & 0.3 & 0.4 \\
			\hline
			bandgap (eV)        & 2.50(3)   & 2.50(7)   & 2.50(7)   & 2.50(3)  & 2.50(7) \\
					\hline
		\end{tabular}
		\label{table2} 
\end{table}

\subsection{Magnetic susceptibility}
\begin{figure}[htbp]
	\centering
	\includegraphics[width=0.9\textwidth]{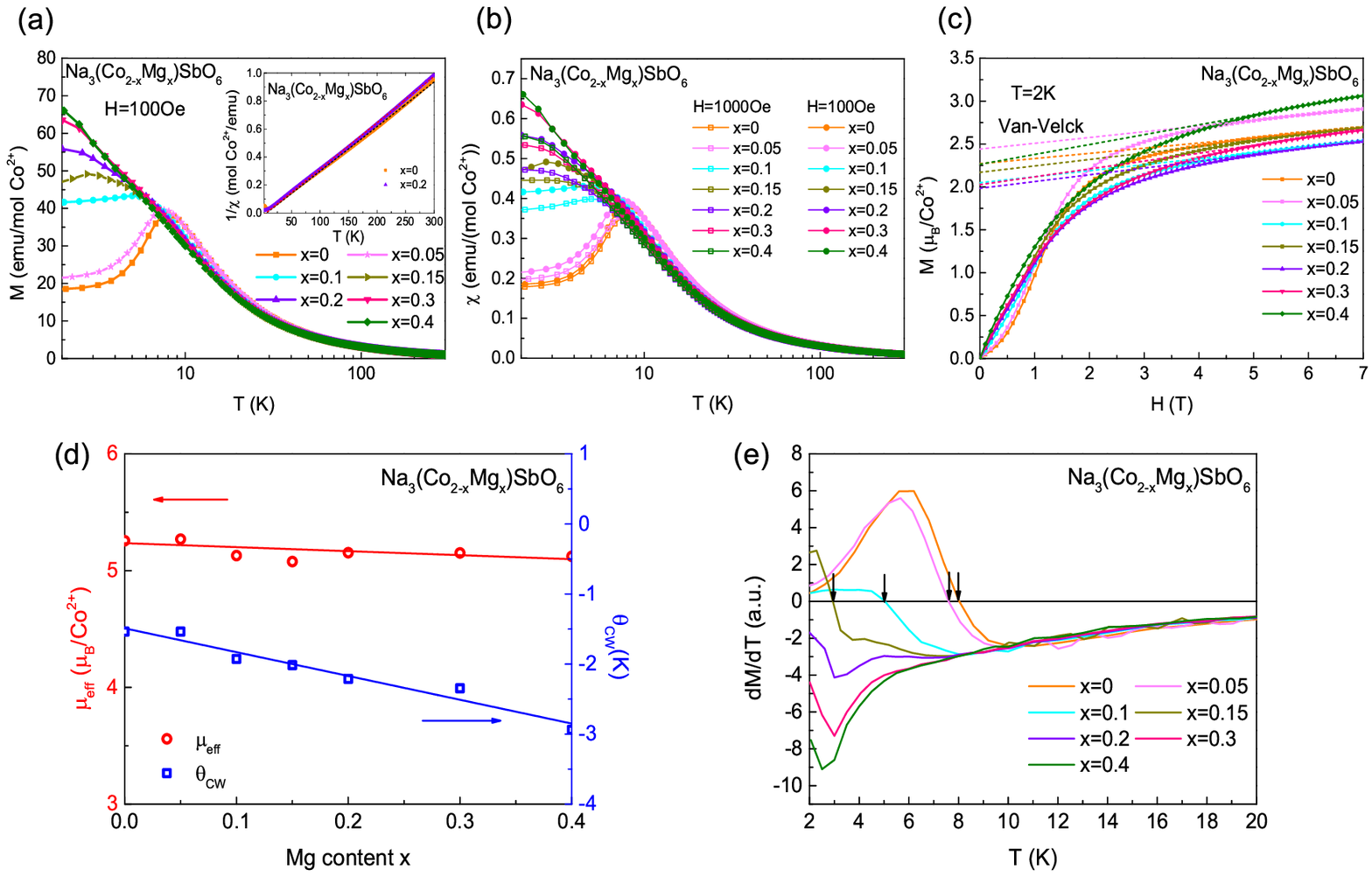}
	%\captionsetup{font=scriptsize}
	\caption{(a) The temperature-dependent DC magnetization of Na$_{3}$(Co$_{2-x}$Mg$_{x}$)SbO$_{6}$ ($x = 0, 0.05, 0.1, 0.15, 0.2, 0.3, 0.4$) measured in field cooling (FC) condition under 100 Oe external field; the inset shows the inverse susceptibility data of two representative Mg concentrations with $x=0$ and $0.2$; the dashed lines are the fits with the Curie-Weiss law.
		(b) The temperature-dependent DC magnetic susceptibility of Na$_{3}$(Co$_{2-x}$Mg$_{x}$)SbO$_{6}$ ($x = 0, 0.05, 0.1, 0.15, 0.2, 0.3, 0.4$) measured in field cooling (FC) condition under 1000 Oe and 100 Oe external field, respectively;
		(c) The iso-thermal magnetization measurement of Na$_{3}$(Co$_{2-x}$Mg$_{x}$)SbO$_{6}$ ($x = 0, 0.05, 0.1, 0.15, 0.2, 0.3, 0.4$) under 2 K. The dash lines indicate the Van Vleck contributions. 
		(d) Extracted effective moment ${\mu_{eff}}$ and Weiss temperature $\theta_{CW}$ as a function of $x$; the solid lines are guides to the eye.
		(e) The first derivative of magnetization versus temperature of Na$_{3}$(Co$_{2-x}$Mg$_{x}$)SbO$_{6}$ ($x = 0, 0.05, 0.1, 0.15, 0.2, 0.3, 0.4$); the arrows mark the positions of $dM/dT=0$ from which the antiferromagnetic transition temperature $T_{N}$ are read.}
	\label{f3}
\end{figure}

In Fig.\ref{f3}, We show the results of DC magnetic measurements of 
Na$_{3}$(Co$_{2-x}$Mg$_{x}$)SbO$_{6}$ ($x = 0, 0.05, 0.1, 0.15, 0.2, 0.3, 0.4$).
We plot the temperature dependence of the magnetization under field cooling 
(FC) condition at $B$$_{ext}$ = 100 Oe in Fig.\ref{f3}(a). Obviously, when the temperature 
is decreasing, there is a sharp maximum around 8 K for the parent compound Na$_{3}$Co$_{2}$SbO$_{6}$.
This corresponds to the onset of an antiferromagnetic ordering at the Neel temperature $T_{N}$ = 8 K. \cite{wong2016zig}
Such sharp maximum shifts to lower temperature with increasing Mg doping. At the same time, this sharp peak deforms wider and flatter. 
Finally, it becomes invisible at $x=0.2$, which means that continuous Mg$^{2+}$ substitution successfully suppresses the antiferromagnetic ordering. Moreover, the magnetization of this system
has a clearly rise approaching the base temperature with increasing Mg contents. Such behavior can also be observed in ${\alpha}$-RuCl$_{3}$ with Ir doping, which may result from the 
uncompensated moments caused by nonmagnetic doping under magnetic order state. \cite{lampen2017destabilization}
In high temperature range, $1/\chi(T)$ can be well 
described by the Curie-Weiss law, $\chi=\chi_{0}+C/(T-\Theta_{CW})$, where
$\chi_{0}$, $C$, and $\Theta_{CW}$ are denoted as the temperature-independent term, Curie-Weiss constant 
and the Weiss temperature, respectively. The inset of Fig.\ref{f3}(a) shows the 
inverse susceptibility data of two representative doping concentrations $x=0$ 
and $0.2$, respectively.
In order to exclude the influences of antiferromagnetic domains on the magnetic susceptibility, we show the DC magnetic measurements of 
Na$_{3}$(Co$_{2-x}$Mg$_{x}$)SbO$_{6}$ ($x = 0, 0.05, 0.1, 0.15, 0.2, 0.3, 0.4$) under field cooling (FC) conditions at $B$$_{ext}$ = 1000 Oe and 100 Oe in Fig.\ref{f3}(b). Comparing with the results at $B$$_{ext}$ = 100 Oe, it can be seen that the magnetic susceptibility at 2 K slightly decreases. However, the antiferromagnetic transition temperature and the transition trend caused by doping are the same in both cases, which can still draw the same conclusions that the antiferromagnetic order is completely suppressed for $x \geq$ 0.2.
In Fig.\ref{f3}(c), we show the iso-thermal magnetization measured at 2 K.
At low magnetic field, $M(H)$ curves gradually increase and approach saturations around 5 T. 
When the external field is larger than 5 T, $M(H)$ curves increase
linearly due to the Van Vleck paramagnetic contribution of Co$^{2+}$, which is similar to other Co-based compounds, such as CsCoCl$_{3}$, Ba$_{8}$CoNb$_{6}$O$_{24}$, etc. \cite{zhou2021comof5, shirata2012experimental, wang2023kmb, rawl2017ba, shiba2003exchange} 
By linear fitting above 5 T, the Van Vleck paramagnetic 
susceptibility $\chi_{vv}$ can be estimated basing on the fitting slope. 
Taking the parent compound Na$_{3}$Co$_{2}$SbO$_{6}$ as an example, 
$\chi_{vv}$ is $\sim$ 0.06035 $\mu_{B}/Co^{2+}$ (= 0.03369 emu mol$^{-1}$ Oe$^{-1}$). In addition, the saturated magnetization 
$M_{s}$ can be obtained by extrapolating the fitted linear curve to the zero field. 
For $x=0$, the resultant $M_{s}$ is about 2.27 $\mu_{B}/Co^{2+}$,
which is in the same level as 1.91 $\mu_{B}/Co^{2+}$ of Ba$_{3}$CoSb$_{2}$O$_{9}$. \cite{shirata2012experimental} 
The fitting slope, the Van Vleck paramagnetic susceptibility $\chi_{vv}$ and the saturated magnetization $M_{s}$ of all samples are tabulated in Table \ref{table3}.
It can be found that the fitting slope and $\chi_{vv}$ monotonically increase with increasing Mg doping levels.
%Interestingly, the variations of $M_{s}$ and $g$ appear a minimum at $x = 0.2$. 

\begin{figure}[htbp]
	\centering
	\includegraphics[width=0.3\textwidth]{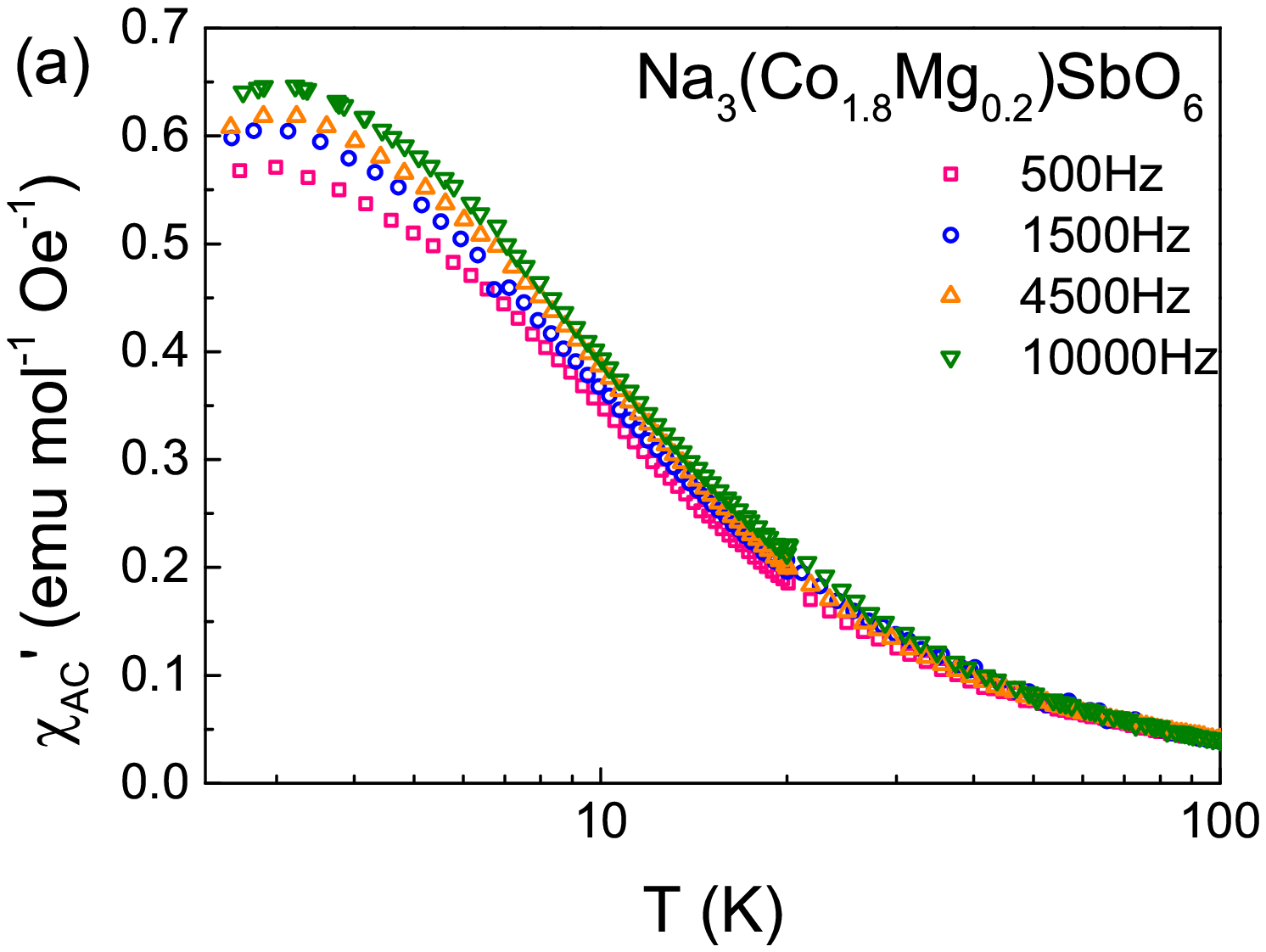}
	\includegraphics[width=0.3\textwidth]{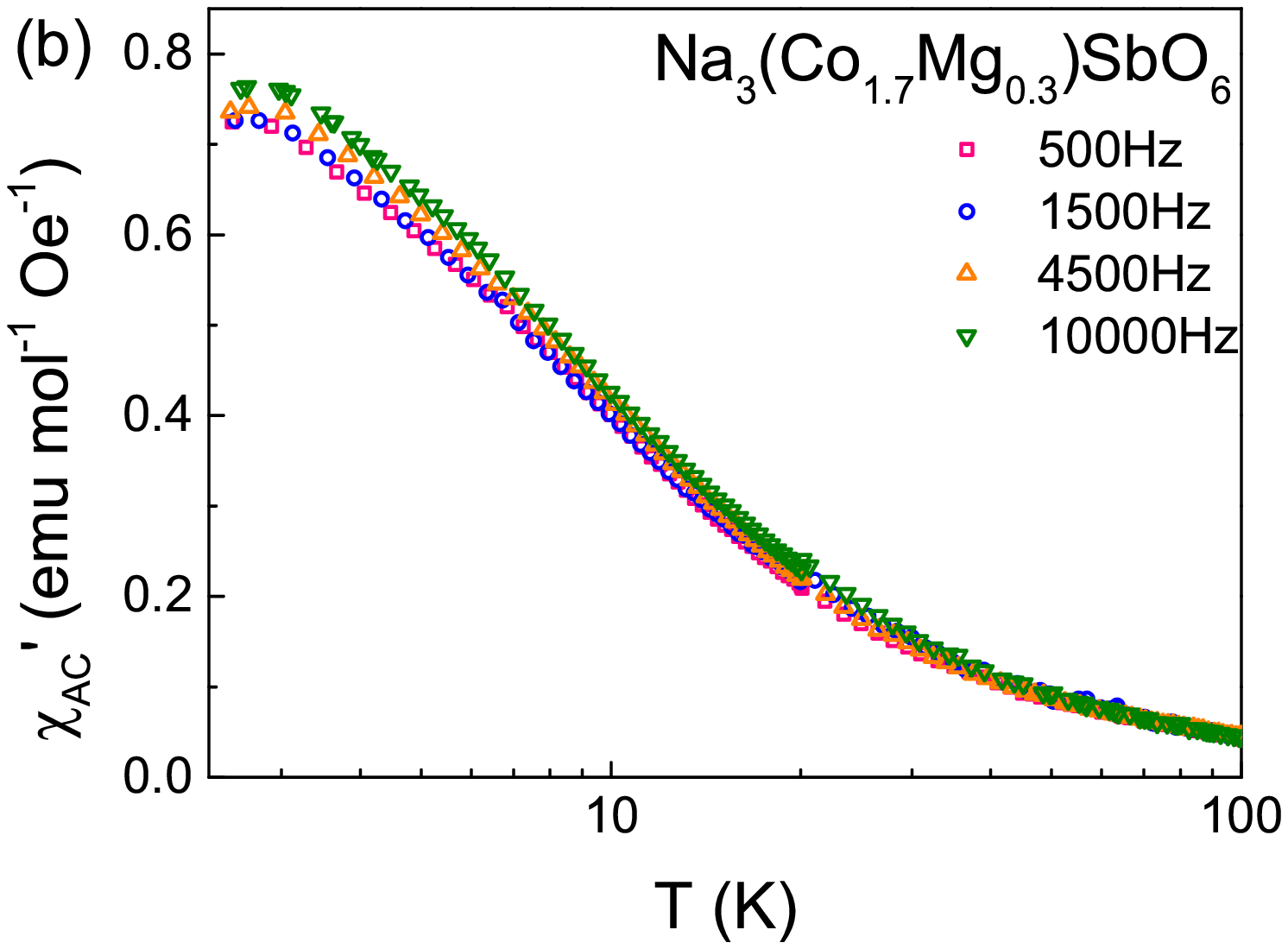}
	\includegraphics[width=0.3\textwidth]{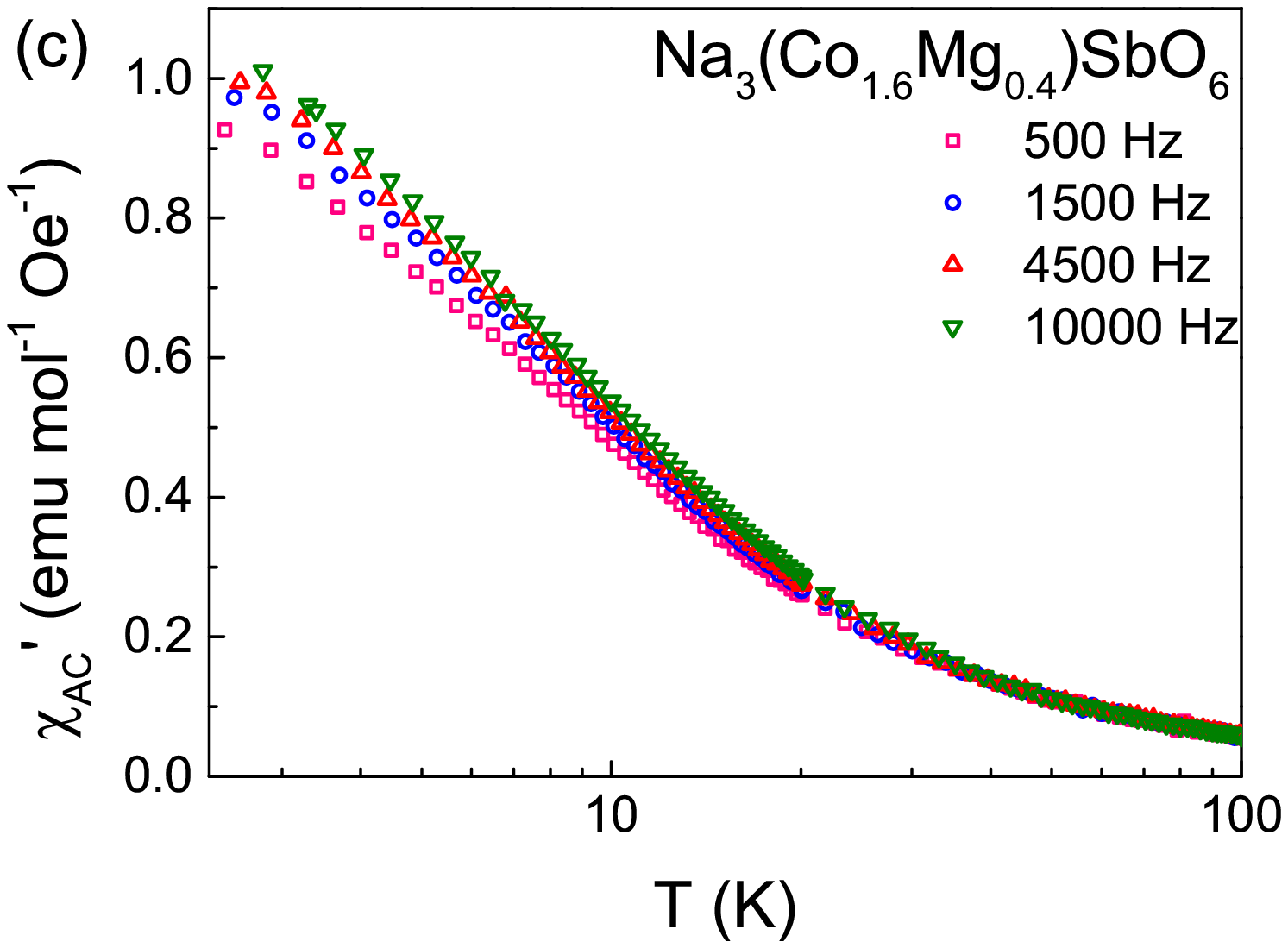}
	\caption{(a)-(c) Temperature dependence of the real part of AC magnetic susceptibility ($\chi^{\prime}_{AC}$) with different driving frequencies under 0.477 Oe external field for $x$ = 0.2, 0.3 and 0.4 compounds, respectively.}
	\label{f4}
\end{figure}

We show the Weiss 
temperatures $\Theta_{CW}$ and the effective magnetic moments ${\mu_{eff}}$ obtained from the fitting of Curie-Weiss law in Fig.\ref{f3}(d). 
The obtained effective moment of Co$^{2+}$ remains constant $\sim$ 5 ${\mu}_B/Co^{2+}$ with increasing $x$. While, for a spin-only $S=3/2$ system, 
the effective moment should be 3.87 ${\mu}_B/Co^{2+}$. Such large practical values of effective moments are 
indicative of the high-spin states with unquenched orbital components 
in current compounds. \cite{fu2023signatures, wong2016zig, yan2019magnetic} 
Similar phenomenon also exists in Na$_{2}$Co$_{2}$TeO$_{6}$. \cite{fu2023suppression} 
By using $J_{eff}$ =1/2 \cite{liu2018pseudospin} and the formula $\mu_{eff} = g\sqrt{J(J+1)}$, a rough estimation of the average magnetic anisotropy of $g$ is $\sim$ 6.07 for the parent compound Na$_{3}$Co$_{2}$SbO$_{6}$. This value is quite large, comparing to the typical spin-only value $g = 2$.
%However, it is proposed that $\mu_{eff}$ is $\sim$ 3.75 ${\mu}_B/Co^{2+}$ expected forthe pseudospin-1/2 case with $g_{J} = -3/2g_{L} + 5/3g_{S}$, where $g_{L,S}$ are the $g$ fator for the effective orbital moment and the spin moment, respetively, given by $g_{L} = -2/3$ and $g_{S} = 2$. \cite{yan2019magnetic}We note that the above experimental value is still lager than this theoretical value.
Unlike the potential quantum behavior induced by substitution in Na$_{2}$Ir$_{1-x}$Ti$_{x}$O$_{3}$, \cite{manni2014effect}
the absolute value of $\Theta_{CW}$ tends to increase slightly with Mg substitutions (from 1.5 K to 2.9 K), implying 
the enhancement of magnetic coupling. In general, the spin-vacancies introduced by 
Mg substitutions will hinder the path of Co-Co super-exchange and weaken the magnetic 
coupling strength. What we observed in current system may be related to multiple competitions between 
complex interactions.
The resultant $\mu_{eff}$, g and $\Theta_{CW}$ of all doping samples are also tabulated in Table \ref{table3}.
In order to distinguish the behaviors of magnetic order transitions in this system 
more clearly, we show the first derivative of magnetization versus temperature
in Fig.\ref{f3}(e). While $dM/dT=0$ corresponds to the extreme point of a $M(T)$ curve, the 
temperatures marked by arrows correspond to the $T_{N}$ values in Fig.\ref{f3}(a). 
We note that as Mg doping increases, the arrow moves towards lower temperature region and disappears for $x \geq 0.2$. It means that 
the antiferromagnetic order in this system is completely suppressed for $x \geq 0.2$, and no long-range magnetic order exists any more.
In general, doping may induce site mixing and cause structural disorder, which can result in the formation of a spin glass state. 
%The spin glass system is associated with short-range order, including two indispensable components like randomness and frustration. \cite{binder1986spin, mydosh1993spin}
In order to test such scenario, we measured the AC magnetic susceptibility of Na$_{3}$(Co$_{2-x}$Mg$_{x}$)SbO$_{6}$ ($x = 0.2, 0.3, 0.4$) samples at several 
different driving frequencies, and show the results in Fig.\ref{f4}(a)-(c).
From the real part of the AC magnetic susceptibility ($\chi^{\prime}_{AC}$), no signatures of spin freezing, frequency dependence or long-range magnetic order can be seen. This indicates that no spin glass ordering is formed in Na$_{3}$(Co$_{2-x}$Mg$_{x}$)SbO$_{6}$ ($x = 0.2, 0.3, 0.4$). 
The broad hump at $\sim$ 3 K can not be attributed to a long-range magnetic
phase transition or spin-glass transition, and similar behavior also occurs in some quantum spin liquid candidates NaYbO$_{2}$ \cite{bordelon2019field} and Na$_{2}$BaCo(PO$_{4}$)$_{2}$. \cite{zhong2019strong}
%Na$_{3}$Co(CO$_{3}$)$_{2}$Cl. \cite{fu2013coexistence}
That is, a NSD state takes place in Na$_{3}$(Co$_{2-x}$Mg$_{x}$)SbO$_{6}$ ($x = 0.2, 0.3, 0.4$).

\begin{table}[htbp]
	\centering
	%\captionsetup{font=scriptsize}
	\caption{The fitting slope, the Van Vleck paramagnetic susceptibility $\chi_{vv}$, the saturated magnetization $M_{s}$, the effective magnetic moment $\mu_{eff}$, the average magnetic anisotropy of $g$ and the Weiss temperature $\Theta_{CW}$ of Na$_{3}$(Co$_{2-x}$Mg$_{x}$)SbO$_{6}$ ($x = 0, 0.05, 0.1, 0.15, 0.2, 0.3, 0.4$).}
	\begin{tabular}{cccccccc}
		\hline
		$x$      & 0   &0.05 & 0.1  &0.15 & 0.2 & 0.3 & 0.4\\
		\hline
		slope                                      & 0.06035    & 0.06734
		& 0.07034   & 0.07483
		& 0.07879   & 0.09157   & 0.11521\\
		$\chi_{vv}$(emu mol$^{-1}$ Oe$^{-1}$)      & 0.03369   & 0.03760
		& 0.03927   & 0.04178
		& 0.04399   & 0.05112 &  0.06432 \\
		$M_{s}$($\mu_{B}/Co^{2+}$)                 & 2.27   & 2.44
		& 2.05   & 2.17
		& 1.98   & 2.03  &  2.26 \\
		$\mu_{eff}$ (${\mu}_B/Co^{2+}$)        & 5.26   & 5.27
		&  5.13   & 5.08
		& 5.15 & 5.15  & 5.13 \\
		$g$                                    & 6.07   & 6.09
		& 5.92    &  5.86
		&  5.95  &  5.95  &   5.92 \\
		$\Theta_{CW}$ (K)                     		 & -1.5   & -1.5
		&  -1.9   & -2.0
		&  -2.2  & -2.3 &  -2.9  \\
		\hline
	\end{tabular}
	\label{table3} 
\end{table} 

\subsection{Specific heat}
\begin{figure}[htbp]
	\centering
	\includegraphics[width=0.3\textwidth]{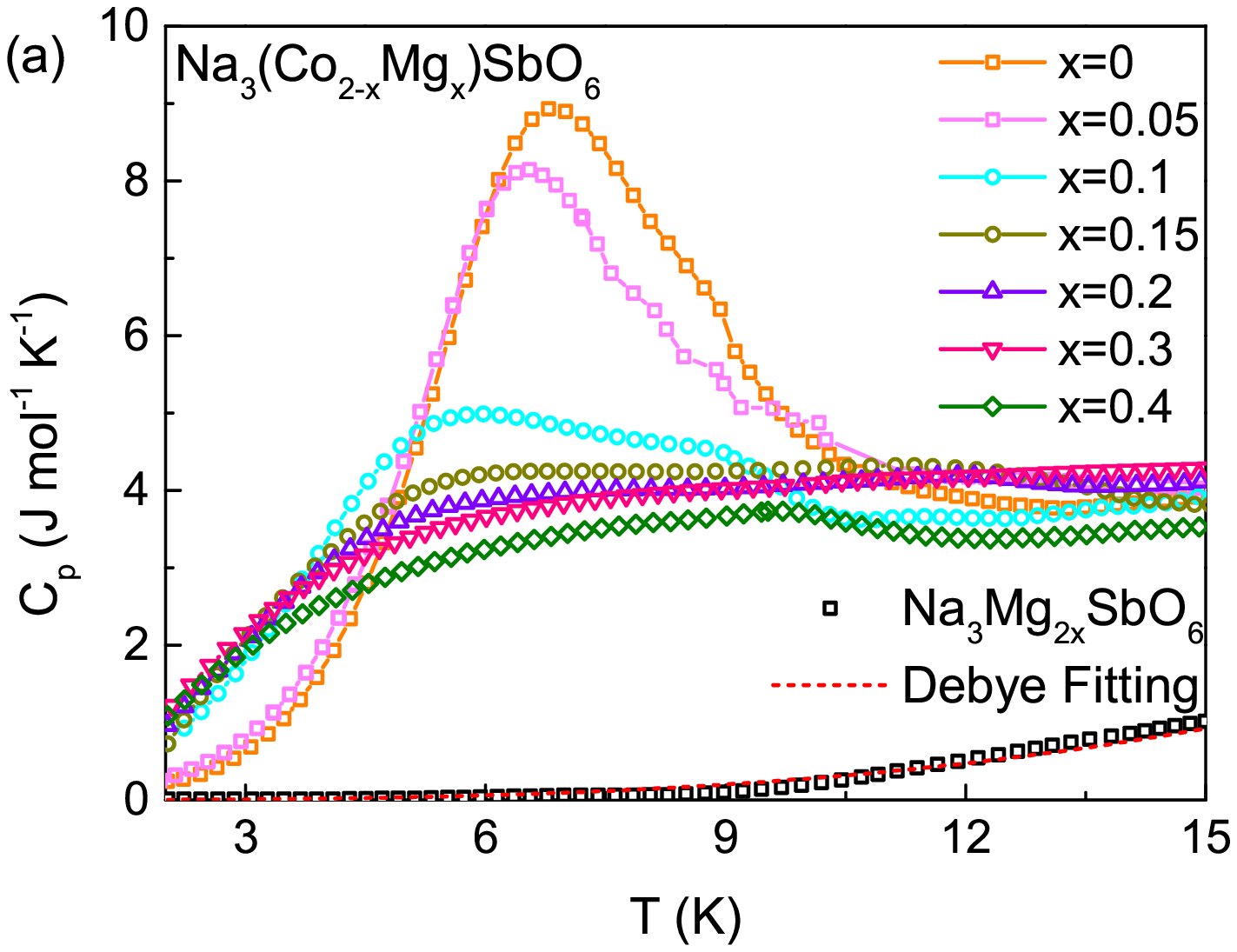}
	\includegraphics[width=0.3\textwidth]{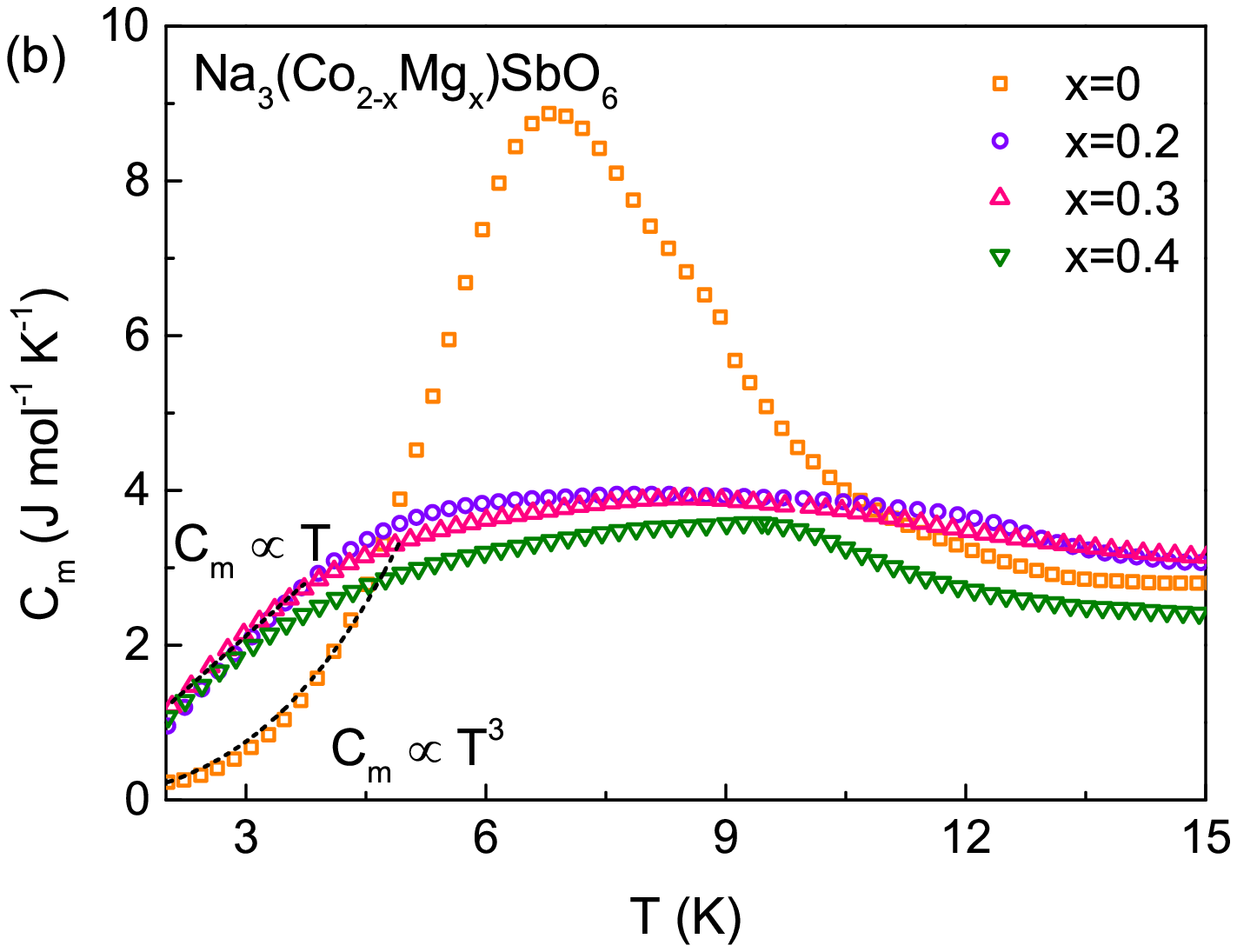}
	\includegraphics[width=0.3\textwidth]{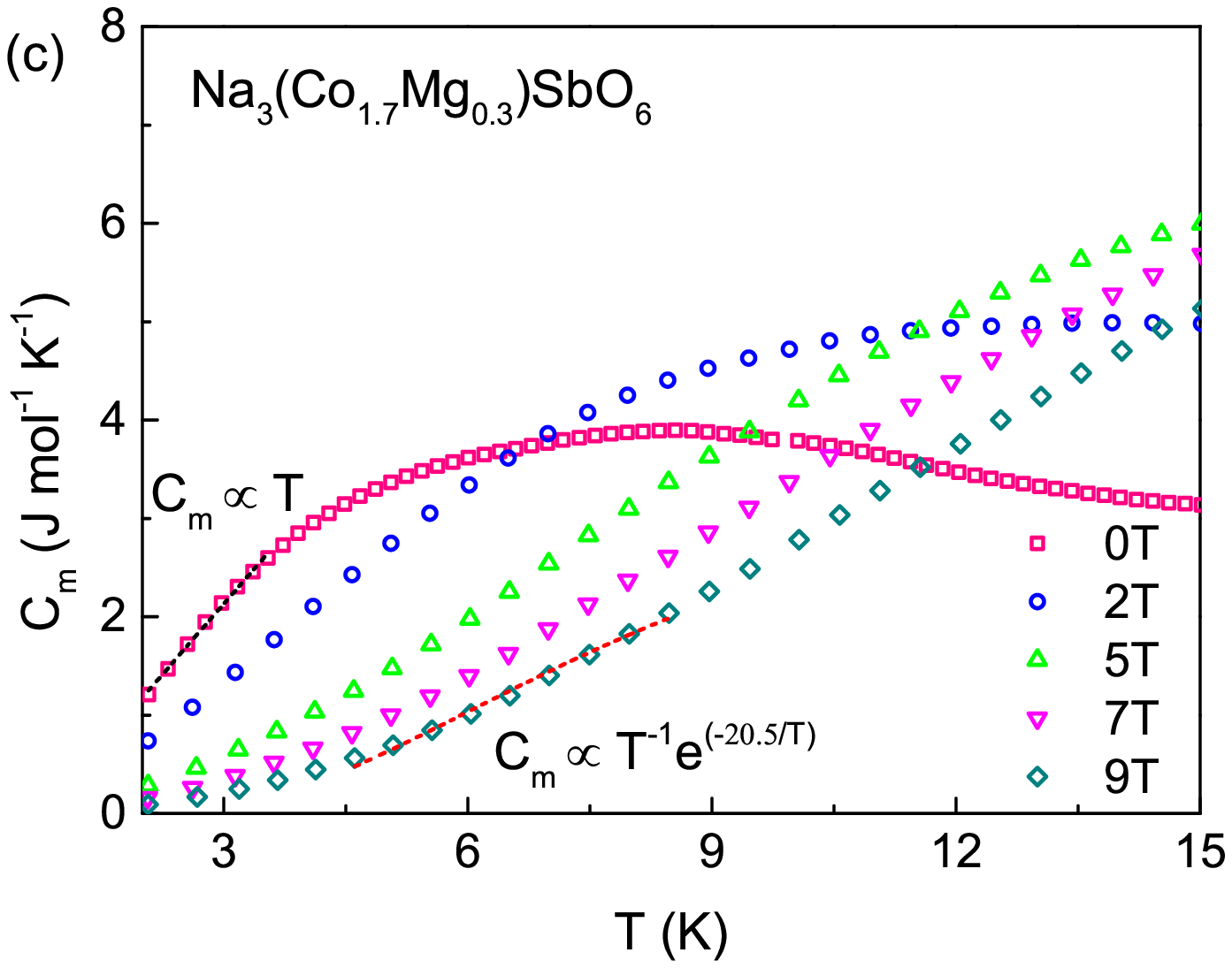}
	\caption{(a) Temperature dependence of specific heat ($C_{p}$) of Na$_{3}$(Co$_{2-x}$Mg$_{x}$)SbO$_{6}$ ($x = 0, 0.05, 0.1, 0.15, 0.2, 0.3, 0.4$) over a temperature range of 2-15 K at zero field. Specific heat of the nonmagnetic reference compound Na$_{3}$Mg$_{2}$SbO$_{6}$ is also shown for comparison, which can be fitted by a Debye model as $C_{p}$ $\propto$ $T^{3}$, denoting by the dashed line.
		(b) Magnetic specific heat ($C_{m}$) of parent compound as well as the NSD compounds with three typical Mg contents $x$ = 0.2, 0.3 and 0.4, respectively; the dash lines are the fitting of compounds with $x$ = 0 and 0.2 at low temperature.
		(c) Magnetic specific heat ($C_{m}$) of Na$_{3}$(Co$_{1.7}$Mg$_{0.3}$)SbO$_{6}$ measured under different magnetic fields. The dash lines at low temperature are the fitting of Na$_{3}$(Co$_{1.7}$Mg$_{0.3}$)SbO$_{6}$ with $B$$_{ext}$ = 0 T and 9 T.}
	\label{f5}
\end{figure}

In order to clearly distinguish the change of ground state caused by Mg doping, we performed specific heat ($C_{p}$) measurements that are the indispensable methods to detect phase transitions and low-energy excitations \cite{zhou2017quantum, wen2019experimental, yamashita2008thermodynamic, yamashita2011gapless, fu2023signatures} on all samples. 
We show $C_{p}(T)$ curves for Na$_{3}$(Co$_{2-x}$Mg$_{x}$)SbO$_{6}$ ($x = 0, 0.05, 0.1, 0.15, 0.2, 0.3, 0.4$) 
and nonmagnetic Na$_{3}$Mg$_{2}$SbO$_{6}$ at zero field in Fig.\ref{f5}(a). Obviously, for parent compound Na$_{3}$Co$_{2}$SbO$_{6}$, 
an $\lambda$-type anomalous peak can be observed around 7 K, indicating 
the onset of the antiferromagnetic phase transition. \cite{wong2016zig, yan2019magnetic}
When Mg doping increases, this anomaly moves to lower temperature region and gradually becomes a broad hump, which is consistent with the change obtained in Fig.\ref{f3}(a). 
For $x \geq 0.2$, these peaks become indistinguishable, and can not be 
identified as a long-range magnetic phase transition. Instead, it is consistent with 
short-range order or no magnetic order at all. \cite{zhou2017quantum, binder1986spin, mydosh1993spin, yamashita2008thermodynamic, yamashita2011gapless, helton2007spin, li2015gapless, ma2018spin}
%However, the AC magnetic susceptibility measurements mentioned above have confirmed the absence of spin glass states in magnetically disordered samples. It may imply that such system enters into a NSD state. 
In Fig.\ref{f5}(b), we show the magnetic specific heat ($C_{m}$) of Na$_{3}$Co$_{2}$SbO$_{6}$ and Na$_{3}$(Co$_{2-x}$Mg$_{x}$)SbO$_{6}$ ($x = 0.2, 0.3, 0.4$) extracted from $C_{p}$.
Because the end-member Na$_{3}$Mg$_{2}$SbO$_{6}$ is well 
described by a Debye phonon heat capacity model  $C_{p} \propto T^{3}$ as shown in Fig.\ref{f5}(a), it can serve as a nonmagnetic analog to estimate 
the phononic contribution of $C_{p}$. 
The specific heat of Na$_{3}$Mg$_{2}$SbO$_{6}$ is defined as $C_{nonmag}$. 
Naturally, $C_{m}$ can be extracted from $C_{p}$ by subtracting $C_{nonmag}$ without the need of any scaling, as done before in $\alpha$-Ru$_{1-x}$Rh$_{x}$Cl$_{3}$. \cite{bastien2022dilution, bastien2019spin}
In low temperature range from 2 K to 5 K, the $C_{m}$ of the parent compound Na$_{3}$Co$_{2}$SbO$_{6}$ can be defined by $\beta$T$^{3}$ with 
$\beta$ = 0.02789 J mol$^{-1}$ K$^{-4}$. It indicates the presence of antiferromagnetic magnons, 
which is consistent with three-dimensional (3D) antiferromagnetic order at low temperature.
Nevertheless, the $C_{m}$ of the NSD samples Na$_{3}$(Co$_{2-x}$Mg$_{x}$)SbO$_{6}$ ($x = 0.2, 0.3, 0.4$) can be well described by $\gamma$T 
in low temperature area. Specifically, the $\gamma$ values are 1.03696 J mol$^{-1}$ K$^{-2}$, 
0.92288 J mol$^{-1}$ K$^{-2}$ and 0.78193 J mol$^{-1}$ K$^{-2}$ for $x$ = 0.2, 0.3 and 0.4, respectively.
Such change of $C_{m}$ from $T^{3}$ to $T$ may imply that magnetic
ground state undergoes a evolution from magnon excitation with $S = 1$ to itinerant quasiparticle excitation with $S = 1/2$. \cite{fu2023signatures}
Interestingly, the non-zero and finite value of $\gamma$ is reminiscent of a gapless 
QSL state with fractional spinon excitations in magnetically disordered systems. \cite{choi2019exotic, helton2007spin, yamashita2011gapless, yamashita2008thermodynamic}

In order to further reveal low-energy excitation behaviors of NSD compounds, we measured  
the magnetic field dependence of $C_{m}$ for Na$_{3}$(Co$_{1.7}$Mg$_{0.3}$)SbO$_{6}$, which is shown in 
Fig.\ref{f5}(c).
The position of the broad hump for $B$$_{ext}$ = 0 T shifts to higher temperature region with increasing magnetic fields, and progressively weakens and eventually fades under $B$$_{ext}$ = 9 T.
Such unique behavior is consistent with reported QSL candidates, such as YbMgGaO$_{4}$, YbZnGaO$_{4}$, etc. \cite{zhong2018field, helton2007spin, li2015gapless, ma2018spin}
Additionally, the suppression of $C_{m}$ at low temperature with magnetic fields implies that the magnetic field has substantial influence on the magnetic ground state.
Comparing with the behavior of $C_{m}$ ($\propto$ T) in zero field, 
it is worth noting that $C_{m}$ can satisfactorily be described 
by a simple model \cite{wolter2017field} under $B$$_{ext}$ = 9 T, which is defined by a formula $C_{m}$ $\propto$ T$^{-1}$exp(-$\Delta$/T), where a bosonic mode 
with gap $\Delta$ and parabolic dispersion in spatial dimensionality $d$ = 2 are used.
The obtained spin gap $\Delta$ is about 20.5(4) K ($\sim$ 1.77 meV).
This variation of $C_{m}$ from a power law behavior to an exponential one with increasing $B$$_{ext}$ may be indicative of an evolution from gapless excitations at zero field to gapped magnons in the fully polarized state. \cite{ding2019gapless, vavilova2023magnetic}

\section{DFT calculations and Discussion}
\begin{figure}[htbp]
	\centering
	\includegraphics[width=0.9\textwidth]{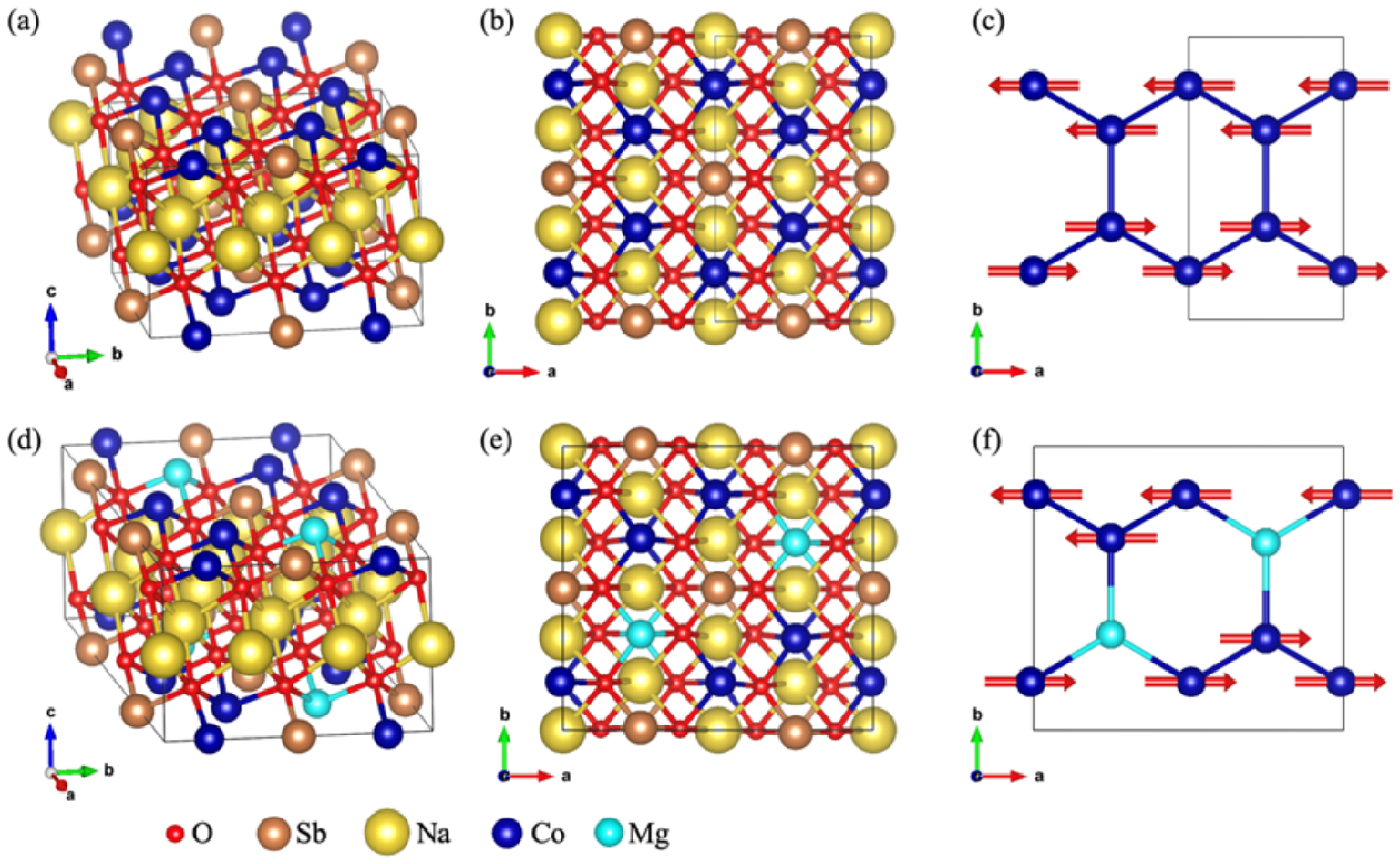}
	\caption{DFT relaxed structure for an undoped system using conventional unit cell (a)-(c) and a doped system using supercell containing eight Co sites with two Co atoms substituted by Mg atoms (d)-(f). (a),(d): Side view; (b),(e): Top view; (c),(f): Magnetic configuration adopted during structural relaxation.}
	\label{f7}
\end{figure}

In addtion to the information about the bond length of Co(Mg)-Co(Mg) and bond-angle of Co(Mg)-O-Co(Mg) in Na$_{3}$(Co$_{2-x}$Mg$_{x}$)SbO$_{6}$ system through the Rietveld refinement, we have also performed density functional theory (DFT) calculations to obtain optimized structures for both pristine and Mg-doped systems. The Mg-doped cell is constructed by substituting two Co atoms for Mg atoms in a supercell containing eight Co sites. Magnetic configuration illustrated in Fig.\ref{f7} (c) and (f) are adopted during the structural relaxation for the pristine system and the doped system, respectively. The resulted structure for the pristine system is shown in Fig.\ref{f7} (a)-(b), where the bond angle of Co-O-Co is $\sim 93.3^{\circ}$, the averaged distance of Co-Co is 3.095 Å (ranging from 3.056 Å to 3.172 Å). The resulted structure for the Mg-doped system is shown in Fig.\ref{f7} (d)-(e), where the bond angle of Co-O-Co is $\sim 91.4^{\circ}$, while the bond angle of Co-O-Mg is $\sim94.6^{\circ}$; the averaged distance of Co(Mg)-Co(Mg) is 3.089 Å (ranging from 2.947 Å to 3.177 Å). As a result, the doping of Mg reduces the distance of Co(Mg)-Co(Mg) and significantly changes the bond angle of Co(Mg)-O-Co(Mg), which must changes the overlap integral and eventually suppress antiferromagnetic ordering.

\begin{figure}[htbp]
	\centering
	\includegraphics[width=0.9\textwidth]{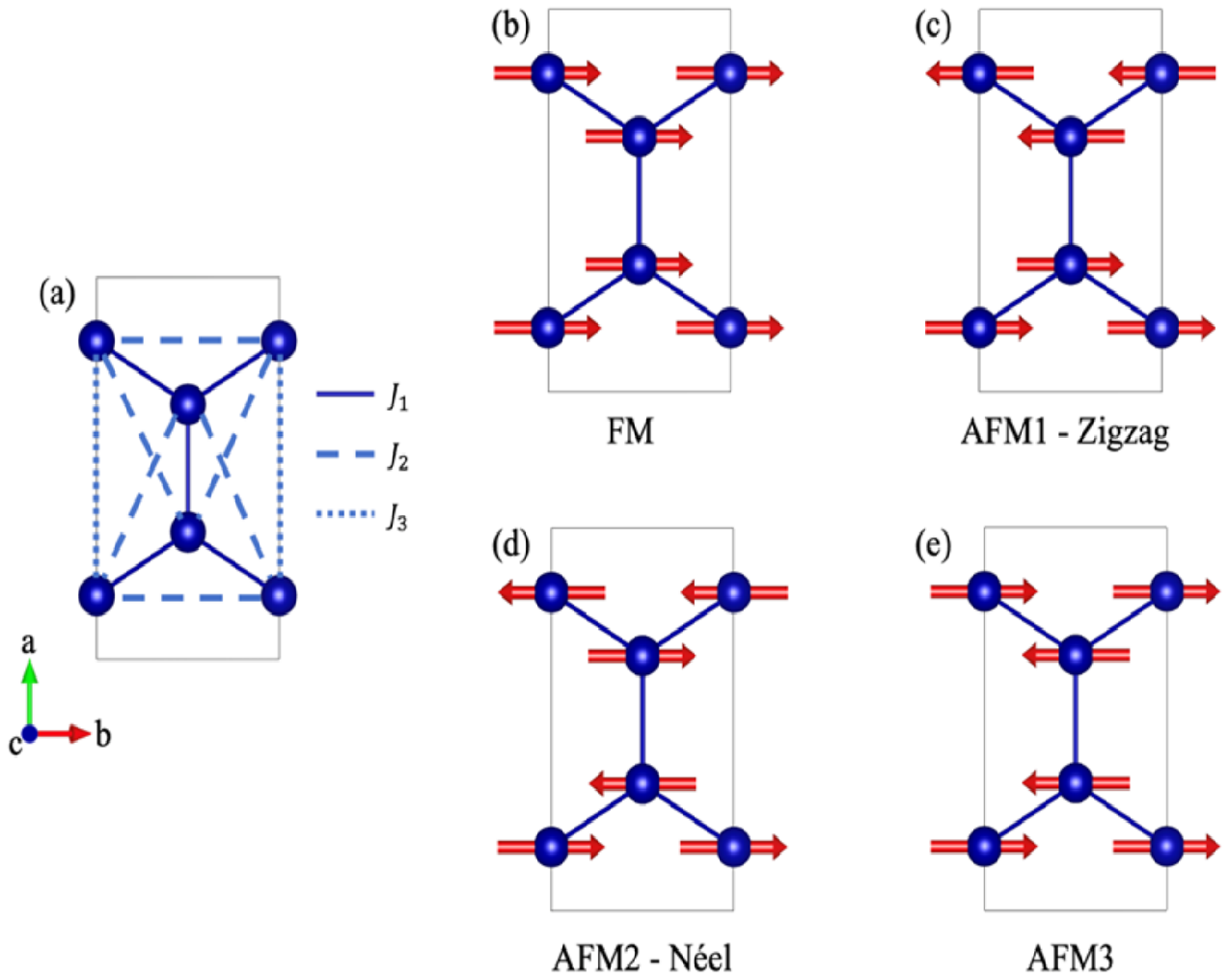}
	\caption{(a) Exchange constants of Heisenberg model for in-plane honeycomb lattice of Co atoms. (b)-(e) Four magnetic structures.}
	\label{f8}
\end{figure}

To evaluate the exchange energies, we have performed DFT calculations using Quantum ESPRESSO (QE). The doping effect was studied using the Virtual Crystal Approximation (VCA). Due to the significant difference in outmost electron configurations between Mg and Co, we utilized Zn and Co instead to construct pseudopotentials, since both Mg and Zn are nonmagnetic dopants. We calculated the total energies for four magnetic configurations (shown in Fig.\ref{f8} (b)-(e)).
Na$_{3}$Co$_{2}$SbO$_{6}$ adopts a layered crystal structure and exhibits an in-plane zig-zag antiferromagnetic (AFM) ordering.\cite{wong2016zig} Given the weak interlayer coupling characteristic of such layered systems, it is reasonable to ignore the inter-plane exchange for simplicity. The in-plane honeycomb arrangement of Co atoms in Na$_{3}$Co$_{2}$SbO$_{6}$ validates a $J_{1} - J_{2} - J_{3}$ Heisenberg model.\cite{oitmaa2011phase} By considering only the nearest-neighbor ($J_{1}$), next-nearest-neighbor ($J_{2}$), and third-nearest-neighbor ($J_{3}$) exchange coupling interactions, the energy differences between the magnetic configurations approximately satisfy the following relationships:
\begin{equation}
	\label{equ:1}
	E^{AFM1} - E^{FM}= E^{AFM1}_{ex} - E^{FM}_{ex} = -4J_{1}-16J_{2}-12J_{3}
\end{equation}
\begin{equation}
	\label{equ:2}
	E^{AFM2} - E^{FM}= E^{AFM2}_{ex} - E^{FM}_{ex} = -12J_{1}-12J_{3}
\end{equation}
\begin{equation}
	\label{equ:3}
	E^{AFM3} - E^{FM}= E^{AFM3}_{ex} - E^{FM}_{ex} = -8J_{1}-16J_{2}
\end{equation}
From the total energy calculations, we derived the values of $J_{1}$, $J_{2}$ and $J_{3}$. The simulation results are summarized in Table \ref{table4}, from which we find that the zig-zag type AFM ordering is primarily governed by $J_{3}$. For the undoped case ($x = 0$), $J_{3}$ is large. However, at doping level of $x = 0.2$, $J_{3}$ is significantly suppressed and all $J_{s}$ become negligibly small. Therefore, the suppression of long-range AFM ordering at doping level of $x = 0.2$ is attributed to the vanishing $J_{3}$ exchange interaction, which well explains the experimental results.
(All structures are relaxed in the zig-zag AFM configuration with atomic forces converged to below 0.003 eV/Å. Total energy calculations was done using a DFT+U method with a Hubbard U parameter of 4.4 eV to account for strong electron correlations.)

\begin{table}[htbp]
	\centering
	%\captionsetup{font=scriptsize}
	\caption{The difference in total energy (eV) per Co between antiferromagnetic (AFM) and ferromagnetic (FM) structures $E^{AFM1} - E^{FM}$, $E^{AFM2} - E^{FM}$ and $E^{AFM3} - E^{FM}$, as well as the solved $J_{1}$, $J_{2}$ and $J_{3}$.}
		\begin{tabular}{cccccccc}
					\hline
			x & $E^{AFM1} - E^{FM}$ & $E^{AFM2} - E^{FM}$ & $E^{AFM3} - E^{FM}$ 
			&  $J_{1}$  &  $J_{2}$  &  $J_{3}$  \\
			\hline
			0 & -0.788    & -0.601
			& 0.095  & -0.070
			& 0.011   & 0.271 \\
			0.2 & -0.007    & -0.059
			& -0.105  & 0.039
			& 0.007   & -0.020 \\
					\hline
		\end{tabular}
		\label{table4} 
\end{table} 

\begin{figure}[htbp]
	\centering
	\includegraphics[width=0.9\textwidth]{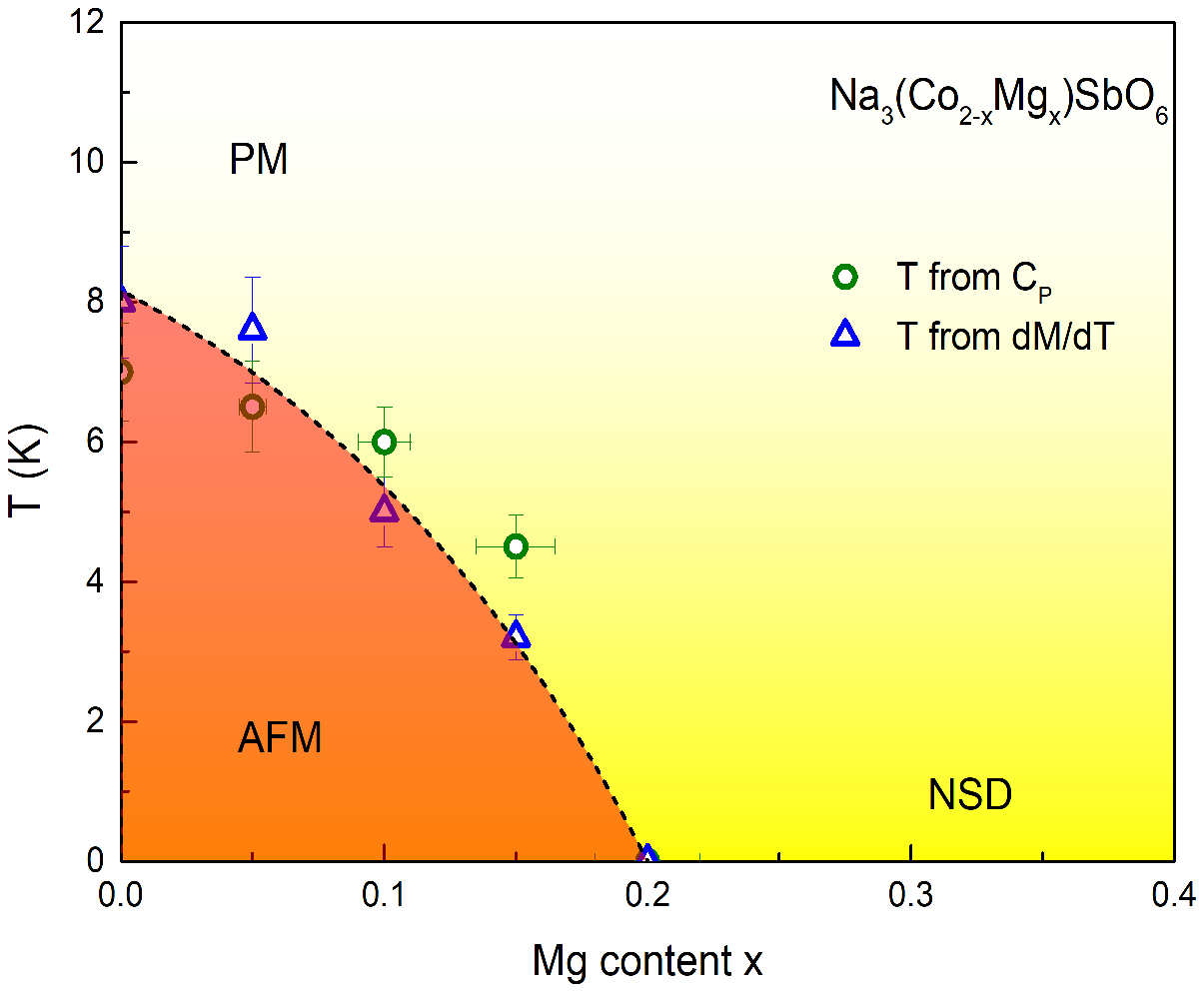}
	\caption{The phase diagram of Na$_{3}$(Co$_{2-x}$Mg$_{x}$)SbO$_{6}$ 
		is consisted of three distinct phases, including the high temperature PM state, low temperature AFM and NSD state. However, due to the lack of characteristic temperature to distinguish PM and NSD states in magnetic and thermodynamic measurements, the phase boundary between them is indistinct, which needs further information.}
	\label{f6}
\end{figure}

Based on the experimental results discussed above, we draw a magnetic phase diagram for Na$_{3}$(Co$_{2-x}$Mg$_{x}$)SbO$_{6}$ in Fig.\ref{f6}. 
Starting from the parent compound ($x = 0$), the antiferromagnetic phase 
transition appears $\sim$ 8 K, forming a frangible zig-zag magnetic order. The transition temperature is gradually 
suppressed with Mg substitutions, and completely disappear at $x$ = 0.2.
The phase boundary temperatures between the antiferromagnetic (AFM) state and paramagnetic (PM) state are determined 
by the differential susceptibility shown in Fig.\ref{f3}(d) and the characteristic 
temperature of $\lambda$-type peak shown in Fig.\ref{f5}(a).
When $x$ is larger than 0.2, this system enters into the 
NSD state, which mimics that a gapless QSL state with fractional spin excitations. Note that, the behaviors of $C_{m}$ in NSD samples ($C_{m}$ $\propto$ T) are different from the parent compound ($C_{m}$ $\propto$ T$^{3}$).

Recently, several studies have discussed the effects of substitutions in Kitaev materials, including 4d$^{5}$ ${\alpha}$-RuCl$_{3}$ 
and 5d$^{5}$ Na$_{2}$IrO$_{3}$. \cite{lampen2017destabilization, do2018short, do2020randomly, manni2014effect, bastien2022dilution}
It is found that the zig-zag type AFM order can effectively be suppressed by doping. Quite differently, ${\alpha}$-(Ru$_{0.8}$Ir$_{0.2}$)Cl$_{3}$ is driven into a magnetically disordered state, but a spin glass state has formed in Na$_{2}$(Ir$_{0.95}$Ti$_{0.05}$)O$_{3}$.
Na$_{3}$(Co$_{2-x}$Mg$_{x}$)SbO$_{6}$ are more similar to the former.
Comparing to ${\alpha}$-(Ru$_{0.8}$Ir$_{0.2}$)Cl$_{3}$, the zig-zag magnetic order in Na$_{3}$(Co$_{2-x}$Mg$_{x}$)SbO$_{6}$ is more frangible and more easily to be driven into a magnetically disorder state, which implies that it is more sensitive to spin vacancies. 
In ${\alpha}$-(Ru$_{0.8}$Ir$_{0.2}$)Cl$_{3}$, the high-energy Majorana fermions and emergent low-energy excitations have been observed, indicating the existence of a QSL-like ground state. \cite{do2020randomly}
Nevertheless, in Na$_{3}$(Co$_{2-x}$Mg$_{x}$)SbO$_{6}$ ($x = 0.2, 0.3, 0.4$), AC magnetic susceptibility measurements exclude the possibility of spin glass state, indicating that Na$_{3}$(Co$_{2-x}$Mg$_{x}$)SbO$_{6}$ system enters into the NSD state with persistent dynamical fluctuations at low temperature.
In ${\alpha}$-RuCl$_{3}$, the honeycomb layers expand, and the interplanar distance
\textit{c$^{\ast}$} is reduced with Ir$^{3+}$ doping. It is equivalent to the case induced by uniaxial pressure along the \textit{c$^{\ast}$} axis, which is predicted to enhance the Kitaev interactions. However, the unit cell is compressed along each crystallographic axis in ${\alpha}$-(Ru$_{1-x}$Rh$_{x}$)Cl$_{3}$. It is comparable to the application of hydrostatic pressure, which has been found to reduce the magnetic ordering temperature in the low-pressure limit. \cite{bastien2022dilution} 
Different from the iso-radius substitutions of Zn$^{2+}$ (r = 0.74Å) for Co$^{2+}$ (r = 0.74Å) in Na$_{3}$Co$_{2}$SbO$_{6}$, the positive chemical pressure in $ab$ plane is induced by Mg$^{2+}$ (r = 0.65Å) doping in our case. It may have the similar mechanism as that of Ir$^{3+}$ or Rh$^{3+}$ doped ${\alpha}$-RuCl$_{3}$, which can make the Na$_{3}$(Co$_{2-x}$Mg$_{x}$)SbO$_{6}$ system enter into the NSD state.

Considering the finite linear term of $C_{m}$ ($\sim$ $\gamma$T) at zero magnetic field, it mimics that a gapless QSL state with fractional quasiparticle excitations, which has been observed in other QSL candidates, 
such as organic salts $\kappa$-(BEDT-TTF)$_{2}$Cu$_{2}$(CN)$_{3}$, EtMe$_{3}$Sb[Pd(dmit)$_{2}$]$_{2}$ and ZnCu$_{3}$(OH)$_{6}$Cl$_{2}$. \cite{zhou2017quantum, broholm2020quantum, yamashita2008thermodynamic, yamashita2011gapless, helton2007spin, choi2019exotic}
Especially, $\gamma$ in Na$_{3}$(Co$_{2-x}$Mg$_{x}$)SbO$_{6}$ ($\gamma$ = 1.03696, 
0.92288 and 0.78193 J mol$^{-1}$ K$^{-2}$ for $x$ = 0.2, 0.3 and 0.4, respectively) are larger than 
those in $\kappa$-(BEDT-TTF)$_{2}$Cu$_{2}$(CN)$_{3}$ ($\gamma$ = 0.020 J mol$^{-1}$ K$^{-2}$), EtMe$_{3}$Sb[Pd(dmit)$_{2}$]$_{2}$ ($\gamma$ = 0.0199 J mol$^{-1}$ K$^{-2}$) and ZnCu$_{3}$(OH)$_{6}$Cl$_{2}$ ($\gamma$ = 0.240 J mol$^{-1}$ K$^{-2}$). It indicates that much larger low energy density of states in 
Na$_{3}$(Co$_{2-x}$Mg$_{x}$)SbO$_{6}$, as $\gamma$ is proportional to the spinon density of states. \cite{yamashita2008thermodynamic}
Comparing to $C_{m}$ $\propto$ $T^{2}$ in ${\alpha}$-(Ru$_{0.8}$Ir$_{0.2}$)Cl$_{3}$ which is the gapless Dirac-like excitations in 2D frustrated lattices, $C_{m}$ is proportional to $T$ in Na$_{3}$(Co$_{2-x}$Mg$_{x}$)SbO$_{6}$ ($x = 0.2, 0.3, 0.4$). Our work is consistent with the RVB
model, where a linear-$T$ dependence of $C_{m}$ at low temperature has been proposed. \cite{do2020randomly, zhou2011spin}
Nevertheless, both models are relative to the fermionic excitations 
expected for a QSL state. \cite{ran2007projected, yamashita2008thermodynamic}

\section{Summary}
To summarize, we report the successful synthesis of Kitaev material Na$_{3}$(Co$_{2-x}$Mg$_{x}$)SbO$_{6}$ ($x = 0, 0.05, 0.1, 0.15, 0.2, 0.3, 0.4$). We investigate both magnetic dilution and chemical pressure effects by substituting Mg$^{2+}$ for Co$^{2+}$  through structural, optical, magnetic and thermodynamic measurements.
No structural transition has been observed, and the bandgaps remain almost constant in all doping levels. 
Basing on magnetic and thermodynamic measurements, we find that the long-range AFM order is gradually suppressed with increasing Mg doping levels, and this system enters into a NSD state at $x \geq 0.2$. Importantly, in NSD samples, $C_{m}$ exhibits the behavior with a finite linear term at low 
temperature in zero field, which is reminiscent of a possible gapless QSL state with fermionic excitations. 
Our investigation indicates that Mg doping is an alternative option to enhance quantum 
fluctuations, providing a potential platform to investigate the NSD state in a Kitaev material.

\section*{Data Availability Statement}
All data generated or analyzed during this study are included in this published article or available from the corresponding author on reasonable request.

\section*{Acknowledgments}
	We thank professor Qingming Zhang for helpful discussions.
	The work at Zhejiang was supported by National Key R\&D Program of China (No. 2022YFA1402701,
	2022YFA1403202), NSF of China (No. 12074333), the Key R\&D Program of Zhejiang Province, China
	(2021C01002).

\section*{Author Contributions}
F.L.N. and J.O.D. conceived this work; J.O.D. conducted the experiments with the help of X.Q.Z., 
L.F.X., X.P., and H.Y.T.; results were analyzed by J.O.D., F.L.N. and X.Q.W; Z.C.X., G.X.Z. and C.C. conducted the calculations. All authors contributed to the 
preparation of this manuscript.

\footnotesize
\itemsep=-3pt plus.2pt minus.2pt 
\bibliographystyle{unsrt}
\bibliography{ref}

\begin{thebibliography}{10}

\bibitem{savary2016quantum}
Lucile Savary and Leon Balents.
\newblock Quantum spin liquids: a review.
\newblock {\em Reports on Progress in Physics}, 80(1):016502, 2016.

\bibitem{li2015rare}
Yuesheng Li, Gang Chen, Wei Tong, Li~Pi, Juanjuan Liu, Zhaorong Yang, Xiaoqun
  Wang, and Qingming Zhang.
\newblock Rare-earth triangular lattice spin liquid: a single-crystal study of
  {Y}b{M}g{G}a{O}$_{4}$.
\newblock {\em Physical Review Letters}, 115(16):167203, 2015.

\bibitem{balents2010spin}
Leon Balents.
\newblock Spin liquids in frustrated magnets.
\newblock {\em Nature}, 464(7286):199--208, 2010.

\bibitem{zhou2017quantum}
Yi~Zhou, Kazushi Kanoda, and Tai-Kai Ng.
\newblock Quantum spin liquid states.
\newblock {\em Reviews of Modern Physics}, 89(2):025003, 2017.

\bibitem{anderson1973resonating}
Philip~W Anderson.
\newblock Resonating valence bonds: {A} new kind of insulator?
\newblock {\em Materials Research Bulletin}, 8(2):153--160, 1973.

\bibitem{kitaev2006anyons}
Alexei Kitaev.
\newblock Anyons in an exactly solved model and beyond.
\newblock {\em Annals of Physics}, 321(1):2--111, 2006.

\bibitem{lyu2022tunneling}
Jing-Jing Lyu, Shi-Qing Jia, and Liang-Jian Zou.
\newblock Tunneling spectra of superconductor-{K}itaev layer-metal junctions.
\newblock {\em Physica B: Condensed Matter}, 625:413483, 2022.

\bibitem{winter2017models}
Stephen~M Winter, Alexander~A Tsirlin, Maria Daghofer, Jeroen van~den Brink,
  Yogesh Singh, Philipp Gegenwart, and Roser Valent{\'\i}.
\newblock Models and materials for generalized {K}itaev magnetism.
\newblock {\em Journal of Physics: Condensed Matter}, 29(49):493002, 2017.

\bibitem{takagi2019concept}
Hidenori Takagi, Tomohiro Takayama, George Jackeli, Giniyat Khaliullin, and
  Stephen~E Nagler.
\newblock Concept and realization of {K}itaev quantum spin liquids.
\newblock {\em Nature Reviews Physics}, 1(4):264--280, 2019.

\bibitem{broholm2020quantum}
C~Broholm, RJ~Cava, SA~Kivelson, DG~Nocera, MR~Norman, and T~Senthil.
\newblock Quantum spin liquids.
\newblock {\em Science}, 367(6475):eaay0668, 2020.

\bibitem{trebst2022kitaev}
Simon Trebst and Ciar{\'a}n Hickey.
\newblock Kitaev materials.
\newblock {\em Physics Reports}, 950:1--37, 2022.

\bibitem{jackeli2009mott}
George Jackeli and Giniyat Khaliullin.
\newblock Mott insulators in the strong spin-orbit coupling limit: from
  {H}eisenberg to a quantum compass and {K}itaev models.
\newblock {\em Physical Review Letters}, 102(1):017205, 2009.

\bibitem{rau2014generic}
Jeffrey~G Rau, Eric Kin-Ho Lee, and Hae-Young Kee.
\newblock Generic spin model for the honeycomb iridates beyond the {K}itaev
  limit.
\newblock {\em Physical Review Letters}, 112(7):077204, 2014.

\bibitem{motome2020hunting}
Yukitoshi Motome and Joji Nasu.
\newblock Hunting {M}ajorana fermions in {K}itaev magnets.
\newblock {\em Journal of the Physical Society of Japan}, 89(1):012002, 2020.

\bibitem{hwan2015direct}
Sae Hwan~Chun, Jong-Woo Kim, Jungho Kim, H~Zheng, Constantinos~C Stoumpos,
  CD~Malliakas, JF~Mitchell, Kavita Mehlawat, Yogesh Singh, Y~Choi, et~al.
\newblock Direct evidence for dominant bond-directional interactions in a
  honeycomb lattice iridate {N}a$_{2}${I}r{O}$_{3}$.
\newblock {\em Nature Physics}, 11(6):462--466, 2015.

\bibitem{singh2012relevance}
Yogesh Singh, S~Manni, J~Reuther, T~Berlijn, R~Thomale, W~Ku, S~Trebst, and
  Philipp Gegenwart.
\newblock Relevance of the {H}eisenberg-{K}itaev model for the honeycomb
  lattice iridates {A}$_{2}${I}r{O}$_{3}$.
\newblock {\em Physical Review Letters}, 108(12):127203, 2012.

\bibitem{sears2015magnetic}
Jennifer~A Sears, M~Songvilay, KW~Plumb, JP~Clancy, Yiming Qiu, Yang Zhao,
  D~Parshall, and Young-June Kim.
\newblock Magnetic order in $\alpha$-{R}u{C}l$_{3}$: {A} honeycomb-lattice
  quantum magnet with strong spin-orbit coupling.
\newblock {\em Physical Review B}, 91(14):144420, 2015.

\bibitem{choi2019spin}
Sungkyun Choi, S~Manni, John Singleton, CV~Topping, Tom Lancaster, SJ~Blundell,
  DT~Adroja, Vivien Zapf, Philipp Gegenwart, and Radu Coldea.
\newblock Spin dynamics and field-induced magnetic phase transition in the
  honeycomb {K}itaev magnet $\alpha$-{L}i$_{2}${I}r{O}$_{3}$.
\newblock {\em Physical Review B}, 99(5):054426, 2019.

\bibitem{banerjee2016proximate}
A~Banerjee, CA~Bridges, J-Q Yan, AA~Aczel, L~Li, MB~Stone, GE~Granroth,
  MD~Lumsden, Y~Yiu, Johannes Knolle, et~al.
\newblock Proximate {K}itaev quantum spin liquid behaviour in a honeycomb
  magnet.
\newblock {\em Nature Materials}, 15(7):733--740, 2016.

\bibitem{banerjee2017neutron}
Arnab Banerjee, Jiaqiang Yan, Johannes Knolle, Craig~A Bridges, Matthew~B
  Stone, Mark~D Lumsden, David~G Mandrus, David~A Tennant, Roderich Moessner,
  and Stephen~E Nagler.
\newblock Neutron scattering in the proximate quantum spin liquid
  $\alpha$-{R}u{C}l$_{3}$.
\newblock {\em Science}, 356(6342):1055--1059, 2017.

\bibitem{do2017majorana}
Seung-Hwan Do, Sang-Youn Park, Junki Yoshitake, Joji Nasu, Yukitoshi Motome,
  Yong~Seung Kwon, DT~Adroja, DJ~Voneshen, Kyoo Kim, T-H Jang, et~al.
\newblock Majorana fermions in the {K}itaev quantum spin system
  $\alpha$-{R}u{C}l$_{3}$.
\newblock {\em Nature Physics}, 13(11):1079--1084, 2017.

\bibitem{zhao2022neutron}
Xiaoxue Zhao, Kejing Ran, Jinghui Wang, Song Bao, Yanyan Shangguan, Zhentao
  Huang, Junbo Liao, Bo~Zhang, Shufan Cheng, Hao Xu, et~al.
\newblock Neutron spectroscopy evidence for a possible magnetic-field-induced
  gapless quantum-spin-liquid phase in a {K}itaev material
  $\alpha$-{R}u{C}l$_{3}$.
\newblock {\em Chinese Physics Letters}, 39(5):057501, 2022.

\bibitem{kim2020proximate}
Beom~Hyun Kim, Shigetoshi Sota, Tomonori Shirakawa, Seiji Yunoki, and Young-Woo
  Son.
\newblock Proximate {K}itaev system for an intermediate magnetic phase in
  in-plane magnetic fields.
\newblock {\em Physical Review B}, 102(14):140402, 2020.

\bibitem{li2021identification}
Han Li, Hao-Kai Zhang, Jiucai Wang, Han-Qing Wu, Yuan Gao, Dai-Wei Qu,
  Zheng-Xin Liu, Shou-Shu Gong, and Wei Li.
\newblock Identification of magnetic interactions and high-field quantum spin
  liquid in $\alpha$-{R}u{C}l$_{3}$.
\newblock {\em Nature Communications}, 12(1):4007, 2021.

\bibitem{zheng2017gapless}
Jiacheng Zheng, Kejing Ran, Tianrun Li, Jinghui Wang, Pengshuai Wang, Bin Liu,
  Zheng-Xin Liu, B~Normand, Jinsheng Wen, and Weiqiang Yu.
\newblock Gapless spin excitations in the field-induced quantum spin liquid
  phase of $\alpha$-{R}u{C}l$_{3}$.
\newblock {\em Physical Review Letters}, 119(22):227208, 2017.

\bibitem{yu2018ultralow}
YJ~Yu, Yang Xu, KJ~Ran, JM~Ni, YY~Huang, JH~Wang, JS~Wen, and SY~Li.
\newblock Ultralow-temperature thermal conductivity of the {K}itaev honeycomb
  magnet $\alpha$-{R}u{C}l$_{3}$ across the field-induced phase transition.
\newblock {\em Physical Review Letters}, 120(6):067202, 2018.

\bibitem{hentrich2018unusual}
Richard Hentrich, Anja~UB Wolter, Xenophon Zotos, Wolfram Brenig, Domenic
  Nowak, Anna Isaeva, Thomas Doert, Arnab Banerjee, Paula Lampen-Kelley,
  David~G Mandrus, et~al.
\newblock Unusual phonon heat transport in $\alpha$-{R}u{C}l$_{3}$: strong
  spin-phonon scattering and field-induced spin gap.
\newblock {\em Physical Review Letters}, 120(11):117204, 2018.

\bibitem{kasahara2018unusual}
Y~Kasahara, K~Sugii, T~Ohnishi, M~Shimozawa, M~Yamashita, N~Kurita, H~Tanaka,
  J~Nasu, Y~Motome, T~Shibauchi, et~al.
\newblock Unusual thermal hall effect in a {K}itaev spin liquid candidate
  $\alpha$-{R}u{C}l$_{3}$.
\newblock {\em Physical Review Letters}, 120(21):217205, 2018.

\bibitem{liu2018pseudospin}
Huimei Liu and Giniyat Khaliullin.
\newblock Pseudospin exchange interactions in $d^{7}$ cobalt compounds:
  {P}ossible realization of the {K}itaev model.
\newblock {\em Physical Review B}, 97(1):014407, 2018.

\bibitem{sano2018kitaev}
Ryoya Sano, Yasuyuki Kato, and Yukitoshi Motome.
\newblock Kitaev-{H}eisenberg {H}amiltonian for high-spin $d^{7}$ {M}ott
  insulators.
\newblock {\em Physical Review B}, 97(1):014408, 2018.

\bibitem{liu2020kitaev}
Huimei Liu, Ji{\v{r}}{\'\i} Chaloupka, and Giniyat Khaliullin.
\newblock Kitaev spin liquid in 3$d$ transition metal compounds.
\newblock {\em Physical Review Letters}, 125(4):047201, 2020.

\bibitem{motome2020materials}
Yukitoshi Motome, Ryoya Sano, Seonghoon Jang, Yusuke Sugita, and Yasuyuki Kato.
\newblock Materials design of {K}itaev spin liquids beyond the
  {J}ackeli--{K}haliullin mechanism.
\newblock {\em Journal of Physics: Condensed Matter}, 32(40):404001, 2020.

\bibitem{vavilova2023magnetic}
E~Vavilova, T~Vasilchikova, A~Vasiliev, D~Mikhailova, V~Nalbandyan, E~Zvereva,
  and SV~Streltsov.
\newblock Magnetic phase diagram and possible {K}itaev-like behavior of the
  honeycomb-lattice antimonate {N}a$_{3}${C}o$_{2}${S}b{O}$_{6}$.
\newblock {\em Physical Review B}, 107(5):054411, 2023.

\bibitem{viciu2007structure}
L~Viciu, Q~Huang, E~Morosan, HW~Zandbergen, NI~Greenbaum, T~McQueen, and
  RJ~Cava.
\newblock Structure and basic magnetic properties of the honeycomb lattice
  compounds {N}a$_{2}${C}o$_{2}${T}e{O}$_{6}$ and
  {N}a$_{3}${C}o$_{2}${S}b{O}$_{6}$.
\newblock {\em Journal of Solid State Chemistry}, 180(3):1060--1067, 2007.

\bibitem{berthelot2012studies}
Romain Berthelot, Whitney Schmidt, AW~Sleight, and MA~Subramanian.
\newblock Studies on solid solutions based on layered honeycomb-ordered phases
  {P}2-{N}a$_{2}${M}$_{2}${T}e{O}$_{6}$ ({M}={C}o, {N}i, {Z}n).
\newblock {\em Journal of Solid State Chemistry}, 196:225--231, 2012.

\bibitem{bera2017spinon}
AK~Bera, B~Lake, FHL Essler, Laurens Vanderstraeten, C~Hubig, Ulrich
  Schollw{\"o}ck, ATMN Islam, A~Schneidewind, and DL~Quintero-Castro.
\newblock Spinon confinement in a quasi-one-dimensional anisotropic
  {H}eisenberg magnet.
\newblock {\em Physical Review B}, 96(5):054423, 2017.

\bibitem{wong2016zig}
Cheryl Wong, Maxim Avdeev, and Chris~D Ling.
\newblock Zig-zag magnetic ordering in honeycomb-layered
  {N}a$_{3}${C}o$_{2}${S}b{O}$_{6}$.
\newblock {\em Journal of Solid State Chemistry}, 243:18--22, 2016.

\bibitem{chamorro2020chemistry}
Juan~R Chamorro, Tyrel~M McQueen, and Thao~T Tran.
\newblock Chemistry of quantum spin liquids.
\newblock {\em Chemical Reviews}, 121(5):2898--2934, 2020.

\bibitem{wen2019experimental}
Jinsheng Wen, Shun-Li Yu, Shiyan Li, Weiqiang Yu, and Jian-Xin Li.
\newblock Experimental identification of quantum spin liquids.
\newblock {\em npj Quantum Materials}, 4(1):12, 2019.

\bibitem{kim2021spin}
Chaebin Kim, Heung-Sik Kim, and Je-Geun Park.
\newblock Spin-orbital entangled state and realization of {K}itaev physics in
  3d cobalt compounds: a progress report.
\newblock {\em Journal of Physics: Condensed Matter}, 34(2):023001, 2021.

\bibitem{lin2021field}
Gaoting Lin, Jaehong Jeong, Chaebin Kim, Yao Wang, Qing Huang, Takatsugu
  Masuda, Shinichiro Asai, Shinichi Itoh, Gerrit G{\"u}nther, Margarita
  Russina, et~al.
\newblock Field-induced quantum spin disordered state in spin-1/2 honeycomb
  magnet {N}a$_{2}${C}o$_{2}${T}e{O}$_{6}$.
\newblock {\em Nature Communications}, 12(1):5559, 2021.

\bibitem{yao2020ferrimagnetism}
Weiliang Yao and Yuan Li.
\newblock Ferrimagnetism and anisotropic phase tunability by magnetic fields in
  {N}a$_{2}${C}o$_{2}${T}e{O}$_{6}$.
\newblock {\em Physical Review B}, 101(8):085120, 2020.

\bibitem{hong2021strongly}
Xiaochen Hong, Matthias Gillig, Richard Hentrich, Weiliang Yao, Vilmos Kocsis,
  Arthur~R Witte, Tino Schreiner, Danny Baumann, Nicol{\'a}s P{\'e}rez, Anja~UB
  Wolter, et~al.
\newblock Strongly scattered phonon heat transport of the candidate {K}itaev
  material {N}a$_{2}${C}o$_{2}${T}e{O}$_{6}$.
\newblock {\em Physical Review B}, 104(14):144426, 2021.

\bibitem{dantas2022disorder}
Vitor Dantas and Eric~C Andrade.
\newblock Disorder, low-energy excitations, and topology in the {K}itaev spin
  liquid.
\newblock {\em Physical Review Letters}, 129(3):037204, 2022.

\bibitem{kao2021vacancy}
Wen-Han Kao, Johannes Knolle, G{\'a}bor~B Hal{\'a}sz, Roderich Moessner, and
  Natalia~B Perkins.
\newblock Vacancy-induced low-energy density of states in the {K}itaev spin
  liquid.
\newblock {\em Physical Review X}, 11(1):011034, 2021.

\bibitem{do2020randomly}
Seung-Hwan Do, CH~Lee, T~Kihara, YS~Choi, Sungwon Yoon, Kangwon Kim, Hyeonsik
  Cheong, Wei-Tin Chen, Fangcheng Chou, H~Nojiri, et~al.
\newblock Randomly {H}opping {M}ajorana {F}ermions in the {D}iluted {K}itaev
  {S}ystem $\alpha$-{R}u$_{0.8}${I}r$_{0.2}${C}l$_{3}$.
\newblock {\em Physical Review Letters}, 124(4):047204, 2020.

\bibitem{fu2023suppression}
Zhongtuo Fu, Ruokai Xu, Song Bao, Yanyan Shangguan, Xin Liu, Zijuan Lu, Yingqi
  Chen, Shuhan Zheng, Yongjun Zhang, Meifeng Liu, et~al.
\newblock Suppression of the antiferromagnetic order by {Z}n doping in a
  possible {K}itaev material {N}a$_{2}${C}o$_{2}${T}e{O}$_{6}$.
\newblock {\em Physical Review Materials}, 7(1):014407, 2023.

\bibitem{fu2023signatures}
Zhongtuo Fu, Ruokai Xu, Yingqi Chen, Song Bao, Hong Du, Jiahua Min, Shuhan
  Zheng, Yongjun Zhang, Meifeng Liu, Xiuzhang Wang, et~al.
\newblock Signatures of a gapless quantum spin liquid in the {K}itaev material
  {N}a$_{3}${C}o$_{2-x}${Z}n$_{x}${S}b{O}$_{6}$.
\newblock {\em Physical Review B}, 107(16):165143, 2023.

\bibitem{yan2019magnetic}
J-Q Yan, Satoshi Okamoto, Yan Wu, Qiang Zheng, HD~Zhou, HB~Cao, and Michael~A
  McGuire.
\newblock Magnetic order in single crystals of
  {N}a$_{3}${C}o$_{2}${S}b{O}$_{6}$ with a honeycomb arrangement of $3d^{7}$
  {C}o$^{2+}$ ions.
\newblock {\em Physical Review Materials}, 3(7):074405, 2019.

\bibitem{johnson2015monoclinic}
Roger~D Johnson, SC~Williams, AA~Haghighirad, John Singleton, Vivien Zapf,
  P~Manuel, II~Mazin, Y~Li, Harald~Olaf Jeschke, R~Valent{\'\i}, et~al.
\newblock Monoclinic crystal structure of $\alpha$-{R}u{C}l$_{3}$ and the
  zigzag antiferromagnetic ground state.
\newblock {\em Physical Review B}, 92(23):235119, 2015.

\bibitem{cao2016low}
Huibo~B Cao, A~Banerjee, J-Q Yan, CA~Bridges, MD~Lumsden, DG~Mandrus,
  DA~Tennant, BC~Chakoumakos, and SE~Nagler.
\newblock Low-temperature crystal and magnetic structure of
  $\alpha$-{R}u{C}l$_{3}$.
\newblock {\em Physical Review B}, 93(13):134423, 2016.

\bibitem{toby2013gsas}
Brian~H Toby and Robert~B Von~Dreele.
\newblock {GSAS-II}: the genesis of a modern open-source all purpose
  crystallography software package.
\newblock {\em Journal of Applied Crystallography}, 46(2):544--549, 2013.

\bibitem{ponosov2024raman}
Yu~S Ponosov, EV~Komleva, EA~Pankrushina, D~Mikhailova, and SV~Streltsov.
\newblock Raman spectroscopy of {N}a$_{3}${C}o$_{2}${S}b{O}$_{6}$.
\newblock {\em JETP Letters}, pages 1--5, 2024.

\bibitem{cairns2022tracking}
Luke~Pritchard Cairns, Ryan Day, Shannon Haley, Nikola Maksimovic, Josue
  Rodriguez, Hossein Taghinejad, John Singleton, and James Analytis.
\newblock Tracking the evolution from isolated dimers to many-body entanglement
  in {N}a{L}u$_{x}${Y}b$_{1-x}${S}e$_{2}$.
\newblock {\em Physical Review B}, 106(2):024404, 2022.

\bibitem{kubelka1931contribution}
Paul Kubelka and Franz Munk.
\newblock A contribution to the optics of pigments.
\newblock {\em Z. Tech. Phys}, 12(593):193, 1931.

\bibitem{tauc1966optical}
J~Tauc, Radu Grigorovici, and Anina Vancu.
\newblock Optical properties and electronic structure of amorphous germanium.
\newblock {\em Physica Status Solidi (B)}, 15(2):627--637, 1966.

\bibitem{li2023investigation}
JW~Ben Li and Brendan~J Kennedy.
\newblock Investigation of hydrogen evolution using
  {N}a$_{3}${M}$_{2}${S}b{O}$_{6}$({M}={C}o$^{2+}$, {N}i$^{2+}$, {C}u$^{2+}$,
  {Z}n$^{2+}$) as photocatalyst.
\newblock {\em Journal of Solid State Chemistry}, 328:124349, 2023.

\bibitem{haussler2022diluting}
Ellen H{\"a}u{\ss}ler, J{\"o}rg Sichelschmidt, Michael Baenitz, Eric~C Andrade,
  Matthias Vojta, and Thomas Doert.
\newblock Diluting a triangular-lattice spin liquid: {S}ynthesis and
  {c}haracterization of {N}a{Y}b$_{1-x}${L}u$_{x}${S}$_{2}$ single crystals.
\newblock {\em Physical Review Materials}, 6(4):046201, 2022.

\bibitem{lampen2017destabilization}
Paige Lampen-Kelley, Arnab Banerjee, Adam~A Aczel, HB~Cao, Matthew~B Stone,
  Craig~A Bridges, J-Q Yan, Stephen~E Nagler, and David Mandrus.
\newblock Destabilization of magnetic order in a dilute {K}itaev spin liquid
  candidate.
\newblock {\em Physical Review Letters}, 119(23):237203, 2017.

\bibitem{zhou2021comof5}
Yadong Zhou, Yanhong Wang, Jiaojiao Cao, Zhuo Zeng, Taiping Zhou, Rongzhen
  Liao, Tao Wang, Zhenxing Wang, Zhengcai Xia, Zhongwen Ouyang, et~al.
\newblock Co{MOF}$_{5}$(pyrazine)({H}$_{2}${O})$_{2}$({M} = {N}b, {T}a):
  two-layered cobalt oxyfluoride antiferromagnets with spin flop transitions.
\newblock {\em Inorganic Chemistry}, 60(17):13309--13319, 2021.

\bibitem{shirata2012experimental}
Yutaka Shirata, Hidekazu Tanaka, Akira Matsuo, and Koichi Kindo.
\newblock Experimental realization of a spin-1/2 triangular-lattice
  {H}eisenberg antiferromagnet.
\newblock {\em Physical Review Letters}, 108(5):057205, 2012.

\bibitem{wang2023kmb}
Yanhong Wang, Shuang Li, Yaling Dou, Hui Li, and Hongcheng Lu.
\newblock {KMB}$_{4}${O}$_{6}${F}$_{3}$({M}= {C}o, {F}e): two-dimensional
  magnetic fluorooxoborates with triangular lattices directed by triangular
  {BO}$_{3}$ units.
\newblock {\em Dalton Transactions}, 52(38):13555--13564, 2023.

\bibitem{rawl2017ba}
R~Rawl, L~Ge, H~Agrawal, Y~Kamiya, CR~Dela Cruz, NP~Butch, XF~Sun, M~Lee,
  ES~Choi, J~Oitmaa, et~al.
\newblock Ba$_{8}${C}o{N}b$_{6}${O}$_{24}$: {A} spin $\frac{1}{2}$
  triangular-lattice {H}eisenberg antiferromagnet in the two-dimensional limit.
\newblock {\em Physical Review B}, 95(6):060412, 2017.

\bibitem{shiba2003exchange}
Hiroyuki Shiba, Yoshifumi Ueda, Kouichi Okunishi, Shojiro Kimura, and Koichi
  Kindo.
\newblock Exchange interaction via crystal-field excited states and its
  importance in {C}s{C}o{C}l$_{3}$.
\newblock {\em Journal of the Physical Society of Japan}, 72(9):2326--2333,
  2003.

\bibitem{manni2014effect}
S~Manni, Yoshifumi Tokiwa, and Philipp Gegenwart.
\newblock Effect of nonmagnetic dilution in the honeycomb-lattice iridates
  {N}a$_{2}${I}r{O}$_{3}$ and {L}i$_{2}${I}r{O}$_{3}$.
\newblock {\em Physical Review B}, 89(24):241102, 2014.

\bibitem{bordelon2019field}
Mitchell~M Bordelon, Eric Kenney, Chunxiao Liu, Tom Hogan, Lorenzo Posthuma,
  Marzieh Kavand, Yuanqi Lyu, Mark Sherwin, Nicholas~P Butch, Craig Brown,
  et~al.
\newblock Field-tunable quantum disordered ground state in the
  triangular-lattice antiferromagnet {N}a{Y}b{O}$_{2}$.
\newblock {\em Nature Physics}, 15(10):1058--1064, 2019.

\bibitem{zhong2019strong}
Ruidan Zhong, Shu Guo, Guangyong Xu, Zhijun Xu, and Robert~J Cava.
\newblock Strong quantum fluctuations in a quantum spin liquid candidate with a
  {C}o-based triangular lattice.
\newblock {\em Proceedings of the National Academy of Sciences},
  116(29):14505--14510, 2019.

\bibitem{yamashita2008thermodynamic}
Satoshi Yamashita, Yasuhiro Nakazawa, Masaharu Oguni, Yugo Oshima, Hiroyuki
  Nojiri, Yasuhiro Shimizu, Kazuya Miyagawa, and Kazushi Kanoda.
\newblock Thermodynamic properties of a spin-1/2 spin-liquid state in a
  $\kappa$-type organic salt.
\newblock {\em Nature Physics}, 4(6):459--462, 2008.

\bibitem{yamashita2011gapless}
Satoshi Yamashita, Takashi Yamamoto, Yasuhiro Nakazawa, Masafumi Tamura, and
  Reizo Kato.
\newblock Gapless spin liquid of an organic triangular compound evidenced by
  thermodynamic measurements.
\newblock {\em Nature Communications}, 2(1):275, 2011.

\bibitem{binder1986spin}
Kurt Binder and A~Peter Young.
\newblock Spin glasses: {E}xperimental facts, theoretical concepts, and open
  questions.
\newblock {\em Reviews of Modern Physics}, 58(4):801, 1986.

\bibitem{mydosh1993spin}
John~A Mydosh.
\newblock {\em Spin glasses: an experimental introduction}.
\newblock CRC Press, 1993.

\bibitem{helton2007spin}
JS~Helton, K~Matan, MP~Shores, EA~Nytko, BM~Bartlett, Y~Yoshida, Y~Takano,
  A~Suslov, Y~Qiu, J-H Chung, et~al.
\newblock Spin dynamics of the spin-1/2 kagome lattice antiferromagnet
  {Z}n{C}u$_{3}$({OH})$_{6}${C}l$_{2}$.
\newblock {\em Physical Review Letters}, 98(10):107204, 2007.

\bibitem{li2015gapless}
Yuesheng Li, Haijun Liao, Zhen Zhang, Shiyan Li, Feng Jin, Langsheng Ling, Lei
  Zhang, Youming Zou, Li~Pi, Zhaorong Yang, et~al.
\newblock Gapless quantum spin liquid ground state in the two-dimensional
  spin-1/2 triangular antiferromagnet {Y}b{M}g{G}a{O}$_{4}$.
\newblock {\em Scientific Reports}, 5(1):16419, 2015.

\bibitem{ma2018spin}
Zhen Ma, Jinghui Wang, Zhao-Yang Dong, Jun Zhang, Shichao Li, Shu-Han Zheng,
  Yunjie Yu, Wei Wang, Liqiang Che, Kejing Ran, et~al.
\newblock Spin-glass ground state in a triangular-lattice compound
  {Y}b{Z}n{G}a{O}$_{4}$.
\newblock {\em Physical Review Letters}, 120(8):087201, 2018.

\bibitem{bastien2022dilution}
Ga{\"e}l Bastien, Ekaterina Vinokurova, Moritz Lange, Kranthi~Kumar Bestha,
  Laura T~Corredor Bohorquez, Gesine Kreutzer, Axel Lubk, Thomas Doert, Bernd
  B{\"u}chner, Anna Isaeva, et~al.
\newblock Dilution of the magnetic lattice in the {K}itaev candidate
  $\alpha$-{R}u{C}l$_{3}$ by {R}h$^{3+}$ doping.
\newblock {\em Physical Review Materials}, 6(11):114403, 2022.

\bibitem{bastien2019spin}
G~Bastien, Maria Roslova, MH~Haghighi, K~Mehlawat, J~Hunger, A~Isaeva, T~Doert,
  M~Vojta, B~B{\"u}chner, and AUB Wolter.
\newblock Spin-glass state and reversed magnetic anisotropy induced by {C}r
  doping in the {K}itaev magnet $\alpha$-{R}u{C}l$_{3}$.
\newblock {\em Physical Review B}, 99(21):214410, 2019.

\bibitem{choi2019exotic}
YS~Choi, CH~Lee, S~Lee, Sungwon Yoon, W-J Lee, J~Park, Anzar Ali, Yogesh Singh,
  Jean-Christophe Orain, Gareoung Kim, et~al.
\newblock Exotic {L}ow-{E}nergy {E}xcitations {E}mergent in the {R}andom
  {K}itaev {M}agnet {C}u$_{2}${I}r{O}$_{3}$.
\newblock {\em Physical Review Letters}, 122(16):167202, 2019.

\bibitem{zhong2018field}
Ruidan Zhong, Mimi Chung, Tai Kong, Loi~T Nguyen, Shiming Lei, and
  Robert~Joseph Cava.
\newblock Field-induced spin-liquid-like state in a magnetic honeycomb lattice.
\newblock {\em Physical Review B}, 98(22):220407, 2018.

\bibitem{wolter2017field}
AUB Wolter, LT~Corredor, L~Janssen, K~Nenkov, Stephan Sch{\"o}necker, S-H Do,
  K-Y Choi, R~Albrecht, J~Hunger, T~Doert, et~al.
\newblock Field-induced quantum criticality in the {K}itaev system
  $\alpha$-{R}u{C}l$_{3}$.
\newblock {\em Physical Review B}, 96(4):041405, 2017.

\bibitem{ding2019gapless}
Lei Ding, Pascal Manuel, Sebastian Bachus, Franziska Gru{\ss}ler, Philipp
  Gegenwart, John Singleton, Roger~D Johnson, Helen~C Walker, Devashibhai~T
  Adroja, Adrian~D Hillier, et~al.
\newblock Gapless spin-liquid state in the structurally disorder-free
  triangular antiferromagnet {N}a{Y}b{O}$_{2}$.
\newblock {\em Physical Review B}, 100(14):144432, 2019.

\bibitem{oitmaa2011phase}
J~Oitmaa and RRP Singh.
\newblock Phase diagram of the {J}$_{1}$-{J}$_{2}$-{J}$_{3}$ {H}eisenberg model
  on the honeycomb lattice: {A} series expansion study.
\newblock {\em Physical Review B—Condensed Matter and Materials Physics},
  84(9):094424, 2011.

\bibitem{do2018short}
Seung-Hwan Do, W-J Lee, S~Lee, YS~Choi, K-J Lee, DI~Gorbunov, J~Wosnitza,
  BJ~Suh, and Kwang-Yong Choi.
\newblock Short-range quasistatic order and critical spin correlations in
  $\alpha$-{R}u$_{1-x}${I}r$_{x}${C}l$_{3}$.
\newblock {\em Physical Review B}, 98(1):014407, 2018.

\bibitem{zhou2011spin}
HD~Zhou, ES~Choi, G~Li, L~Balicas, CR~Wiebe, Yiming Qiu, JRD Copley, and
  Jason~S Gardner.
\newblock Spin liquid state in the {S}= 1/2 triangular lattice
  {B}a$_{3}${C}u{S}b$_{2}${O}$_{9}$.
\newblock {\em Physical Review Letters}, 106(14):147204, 2011.

\bibitem{ran2007projected}
Ying Ran, Michael Hermele, Patrick~A Lee, and Xiao-Gang Wen.
\newblock Projected-{W}ave-{F}unction {S}tudy of the {S}pin-1/2 {H}eisenberg
  {M}odel on the {K}agom{\'e} {L}attice.
\newblock {\em Physical Review Letters}, 98(11):117205, 2007.

\end{thebibliography}

\end{document}